\documentstyle[12pt,epsfig]{article}
\begin{document}
\sloppy
\thispagestyle{empty}

\begin{flushright}
\end{flushright}
\vspace*{\fill}
\begin{center}
{\LARGE\bf  Transverse Spin  Effects In Diffractive Hadron
Leptoproduction}\\
\vspace{2em}
\large
S.V.Goloskokov
\footnote{Email: goloskkv@thsun1.jinr.ru}
\\
\vspace{2em}
Bogoliubov Laboratory of Theoretical  Physics,\\
  Joint Institute for Nuclear Research.
 \\
 Dubna 141980, Moscow region, Russia.
\end{center}
\vspace*{\fill} %
\begin{abstract}
\noindent We  consider  double spin asymmetries for longitudinally
polarized leptons and transversely polarized protons in
diffractive vector meson and $Q \bar Q$ production at high
energies  within the two-gluon model. The asymmetry predicted for
meson production is quite small. The $A_{lT}$ asymmetry for $Q
\bar Q$ production contains  two independent terms which are large
and can be used to  obtain information on the polarized gluon
distributions in the proton.
\end{abstract}

\vspace*{\fill}
\newpage
%


\section{Introduction} \label{sect1}

Study of  the hadron structure  is a fundamental problem of modern
physics. One of the important objects here is  parton
distributions in a nucleon. The cross section of inclusive hadron
production is expressed in terms of ordinary parton distributions
where partons have the same momenta. More general structures in a
nucleon can be studied in deeply virtual Compton scattering or in
diffractive hadron leptoproduction. Really, the kinematics of
there processes requires a nonzero longitudinal momentum $\zeta p$
carried by a two-parton system. As a result, the parton momenta
cannot be equal, and such reactions are expressed in terms of
skewed parton distributions (SPD) \cite{rad,ji}. The factorization
of the amplitude of diffractive hadron production into a hard
subprocess and a soft proton matrix element-- SPD has been shown
in \cite{coll}. The diffractive charm $Q \bar Q$ production and
$J/\Psi$ production are determined by the gluon SPD ${\cal
F}_\zeta(x)$, because the charm  component in the proton is small.
The processes with  light quarks are predominated at small Bjorken
$x \leq 0.1$  by the Pomeron exchange which can be associated with
a two--gluon state \cite{low}. Both quark and gluon SPD will
contribute here for $x>0.1$.

Sensitivity of diffractive lepto and photoproduction to the gluon
density in the proton gives an excellent tool to test these
structure functions. Intensive experimental study of diffractive
processes were performed in DESY (see e.g.
\cite{zeus97,jpsi1,h1_99,dijet} and references therein). The
longitudinal double spin asymmetry in vector meson production has
been analysed in \cite{hermes}. Theoretical investigation of the
diffractive vector meson production was conducted on the basis of
different models where sensitivity of the experimental observables
to polarized parton distributions was studied. Within the two-
gluon exchange model, the typical scale $\bar Q^2=(Q^2+M_V^2)/4$
was found for vector mesons production \cite{rys,J-Psi}. The cross
sections of light and heavy meson production plotted versus this
variable looks similar \cite{clebaux}.  In papers
\cite{rys1,cud,ivan}   the cross section for longitudinal and
transverse photon polarization was analyzed. It was shown that a
longitudinally polarized photon gives a predominat contribution to
the cross section for $Q^2 \to \infty$. The cross section with
transverse photon polarization is suppressed as a power of $Q$.
Investigation of the vector meson production within the SPD
approach was performed by many authors (see e.g
\cite{mank98,hua_kr}). Within the SPS approach, one can
 study simultaneously the imaginary and real parts
of diffractive amplitudes. In paper \cite{mank}, the double spin
asymmetry for longitudinal photon and proton polarization in
$J/\Psi$ production was estimated. The contribution with
transversely polarized photons and a vector meson is important for
the spin observable like $A_{ll}$ asymmetry. Unfortunately, for
the light meson production, this higher twist transition amplitude
is not well defined because of the present infrared singularities
\cite{mank00}.

Another possibility to study SPD in the proton is based on the
quark pair leptoproduction. Theoretical analysis of  the
diffractive $Q \bar Q$ production which can be observed as two jet
events in lepton--proton interaction was carried out e.g. in
\cite{die95, bart96,  ryskin97, schaef}. It was shown that the
cross sections of diffractive quark- antiquark production are
expressed in terms of gluon distributions as in the case of vector
meson production. However, the scale variable in the structure
function is here determined by the transverse momentum of a
produced quark. Spin effects in diffractively produced
quark-antiquark pair for longitudinally polarized lepton and
proton was discussed in \cite{bart99} where the diffractive
contribution to the $g_1$ structure function was calculated.

Thus, the diffractive reactions should play a key role in the
study of the gluon structure of the proton at small $x$. In the
case of polarized particles, the spin-dependent gluon
distributions can be investigated. Previously, spin asymmetries
for longitudinally polarized particles were mainly analyzed. In
future, it will be an excellent possibility to study spin effects
with transversely polarized target at HERMES. Numerous proposals
for possible experiments with this target were discussed in
\cite{transverse}. Such experiments should shed light on the
polarized parton distributions which are responsible on the
transverse spin effects in the hadron.

In this paper we consider  double spin asymmetries for
longitudinally polarized leptons and transversely polarized
protons in diffractive vector meson and $Q \bar Q$ production at
high energies. Some preliminary results in this field were
published in \cite{golos01}. The two-gluon exchange model with the
spin-dependent $gg$-proton coupling is used. This means that our
results should be applicable for reactions with  heavy quarks
which are determined by the gluons exchange. For processed with
light quarks our predictions should be valid at small $x$ ($x \le
0.1$ e.g.). The cross section of hadrons leptoproduction can be
decomposed into the leptonic and hadronic tensors and the
amplitude of hadron production through the $\gamma^\star gg$
transition to the vector meson or $Q \bar Q$ states. After
describing the kinematics of the process in Sect. 2, we analyze
the structure of the leptonic and hadronic tensors in Sect. 3.  In
Sect. 4, we calculate the polarized cross section of vector meson
leptoproduction.  Connection of the two-gluon approach with skewed
gluon distributions is discussed, too. Similar results for
diffractive $Q \bar Q$ production are presented in Sect. 5. The
numerical results for the diffractive vector meson and  production
at HERMES and COMPASS energies and our prediction for the $A_{lT}$
asymmetry can be found in Sects 6 and 7. We finish with concluding
remarks in Sect. 8.

\section{Kinematics of Diffractive Hadron Leptoproduction}
\label{sect2}
Let us study the diffractive hadron production in
lepton-proton reactions
\begin{equation}
\label{react} l+p \to l+p +H
\end{equation}
at high energies in a lepton-proton system. The hadron state $H$
in this reaction can contain a vector meson or a $Q \bar Q$ system
which can be detected as two final jets. The reaction
(\ref{react}) can be described in terms of the kinematic variables
which are defined as follows:
\begin{eqnarray}
\label{momen} q^2= (l-l')^2=-Q^2,\;t=r_P^2=(p-p')^2, \nonumber
\\  y=\frac{p \cdot q}{l  \cdot p},\;x=\frac{Q^2}{2p  \cdot q},\;
x_P=\frac{q \cdot (p-p')}{q \cdot p},\; \beta=\frac{x}{x_P},
\end{eqnarray}
where $l, l'$ and $p, p'$ are the initial and final lepton and
proton momenta, respectively, $Q^2$ is the photon virtuality, and
$r_P$ is the momentum carried by the Pomeron. The variable $\beta$
is used in $Q \bar Q$ production. In this case  the effective mass
of a produced quark system is equal to $M_X^2=(q+r_P)^2$ and can
be quite large. The new variable $\beta=x/x_P \sim
Q^2/(M_X^2+Q^2)$ which appears in this case can vary from 0 to 1.
For diffractive vector meson production, $M_X^2=M_V^2$ and $\beta
\sim 1$ for large $Q^2$. From the mass-shell equation for the
vector--meson momentum $K_V^2=(q+r_P)^2=M_V^2$, we find that for
these reactions
\begin{equation}
\label{x_P}
x_P \sim \frac{m_V^2+Q^2+|t|}{s y}
\end{equation}
and is small at high energies. This variable is not fixed for $Q
\bar Q$ production.

We use the light--cone variables that are determined as $a_\pm=a_0
\pm a_z$. In these variables, the scalar production of two
4-vectors looks like
$$ a \cdot
b=\frac{1}{2}(a_{+} b_{-} +a_{-} b_{+}) - \vec a_\bot \vec
b_\bot,$$ where $\vec a_\bot$ and $\vec b_\bot$ represent the
transverse parts of the momenta. In calculation, the center of
mass system is used where the momenta of the initial  lepton and
proton are going along the $z$ axis and have the form
\begin{equation}
l=(p_{+},\frac{\mu^2}{p_{+}},\vec 0),\quad p=(\frac{m^2}{p_{+}},p_{+},\vec
0).
\end{equation}
Here $\mu$ and $m$ are the lepton and proton mass. The energy of
the lepton--proton system then reads as $  s \sim p_{+}^2$.

The momenta are carried by the photon and the Pomeron and can be
written as follows:
\begin{eqnarray}
q&=&(y p_{+},-\frac{Q^2}{p_{+}},\vec q_\bot),\quad
|q_\bot|=\sqrt{Q^2 (1-y)};
\nonumber \\
r_P&=&(-\frac{|t|}{p_{+}},x_P p_{+},\vec r_\bot),\;\;
|r_\bot|=\sqrt{|t| (1-x_P)}.
\end{eqnarray}
We can determine the spin vectors with positive helicity of the
lepton and the proton by
\begin{eqnarray}
\label{sp}
s_l&=&\frac{1}{\mu}(p_{+},-\frac{\mu^2}{p_{+}},\vec 0),\quad\;
s_l^2=-1,\quad  s_l \cdot l=0;\nonumber \\
s_p&=&\frac{1}{m}(\frac{m^2}{p_{+}},-p_{+},\vec 0),\quad
s_p^2=-1,\quad  s_p \cdot p=0.
\end{eqnarray}
The polarization vector for a transversely polarized  target can
be written in the form
\begin{equation}\label{st}
s_p^\perp=(0,0,\vec s_\perp),\quad \vec s_\perp^2=1.
\end{equation}

\section{Structure of Leptonic and Hadronic Tensors}\label{sect3}
\subsection{Leptonic Tensor}
 The
structure of the leptonic tensor is quite simple \cite{efrem},
because the lepton is a point-like object
\begin{eqnarray}
\label{lept}
{\cal L}^{\mu \nu}(s_l)&=& \sum_{spin \ s_f} \bar u(l',s_f)
\gamma^{\mu} u(l,s_l) \bar u(l,s_l) \gamma^{\nu}
u(l',s_f)
\nonumber \\
&=& {\rm Tr} \left [  (/\hspace{-1.7mm} l+\mu) \frac{1+
\gamma_5 /\hspace{-2.1mm} s_l}{2}
\gamma^{\nu} ( /\hspace{-1.7mm} l'+\mu) \gamma^{\mu}
 \right] .
\end{eqnarray}
Here $l$ and $l'$ are the initial and final lepton momenta, and $s_l$
is a spin vector of the initial lepton determined in (\ref{sp}).

The sum and difference of the cross sections with parallel and
antiparallel longitudinal polarization of a proton and a lepton
are expressed in terms of the spin-average and spin--dependent
hadron and  lepton tensors. The latter is determined  by the
relation
\begin{equation}
\label{l+-}
{\cal L}^{\mu \nu}(\pm)=\frac{1}{2}({\cal L}^{\mu \nu}(+\frac{1}{2})
\pm {\cal L}^{\mu \nu}(-\frac{1}{2})),
\end{equation}
where ${\cal L}^{\mu;\nu}(\pm\frac{1}{2})$ are the tensors with
helicity of the initial  lepton equal to $\pm 1/2$. The tensors
(\ref{l+-}) look like
\begin{eqnarray}
\label{lpm}
{\cal L}^{\mu \nu}(+)&=& 2 (g^{\mu \nu} l \cdot q + 2 l^\mu l^\nu -
l^\mu q^\nu - l^\nu q^\mu),\nonumber \\
{\cal L}^{\mu \nu}(-)&=& 2 i \mu \epsilon^{\mu\nu\delta\rho} q_{\delta}
(s_{l})_{\rho}.
\end{eqnarray}

\subsection{Proton Two Gluon Coupling and Hadron Tensor}

The $Q \bar Q$ system which appears in the final state or passes
into the vector meson in reaction (\ref{react})  can be produced
in two ways. The first one is the photon interaction with the $Q
\bar Q$ state from the proton. This contribution can be connected
with the quark distribution in a nucleon. The other contribution
is determined by the photon-gluon fusion which produces $Q \bar Q$
system. The quark pair must be in a color singlet state to produce
the vector meson. This means that the gluon state should be
colorless too and contains two gluons at least. We are working in
the low $x$ region where the gluon contribution predominates. It
is associated with the Pomeron that describes diffractive
processes at high energies. In QCD--inspired models, the Pomeron
is usually represented as a two-gluon object.

Properties of the gluon structure functions are determined by the
nonperturbative effects inside the proton. We shall analyze only
the matrix structure of two-gluon coupling with the proton within
the quark-diquark model \cite{kroll} where the proton is composed
of a quark and a diquark. Composite scalar and vector diquarks
provide an effective description of nonperturbative effects in the
gluon-proton interaction. The vector diguark produces spin-flip
effects in the proton coupling with the gluon.
\begin{figure}
\epsfxsize=7cm \centerline{\epsfbox{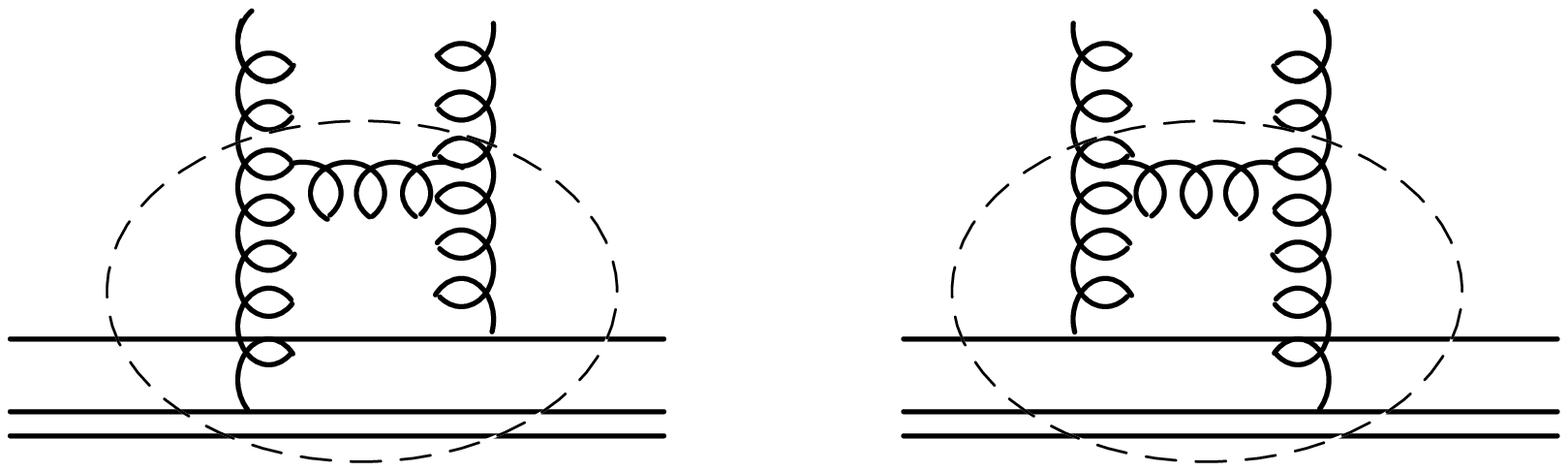}} \caption{Graphs
which give the leading Log contribution to the $gg p$ vertex in
$\alpha^2$ order.}
\label{llog}       
\end{figure}

It has been shown in \cite{grib77} that the leading contribution
like $\alpha_s \left[ \alpha_s \ln{(1/x)}\right]^n$ to the Pomeron
is determined by the gluon ladder graphs. In the $\left[ \alpha_s
\right]^2$ order we have in the model two ladder graphs shown in
Fig. \ref{llog} with $\alpha_s^2 \ln{(1/x)}$ behavior. We include
to the gluons coupling with the proton the gluon ladder, except
two upper $t$-channel gluons in Fig. \ref{llog}. This coupling is
shown in the graphs of Fig. \ref{llog} by the blob. In what
follows, we shall calculate the imaginary part of the Pomeron
contribution to the scattering amplitude which dominates in the
high-energy region. This contribution is equivalent to the
$t$--channel cut in the gluon--loop graphs. In the diquark model,
the following structures in the coupling appears
\begin{eqnarray}\label{ver}
V_{pgg}^{\alpha\beta}(p,t,x_P,l_\perp)&=& B(t,x_P,l_\perp)
(\gamma^{\alpha} p^{\beta} + \gamma^{\beta} p^{\alpha})+\frac{i
K(t,x_P,l_\perp)}{2 m} (p^{\alpha} \sigma ^{\beta
\gamma} r_{\gamma} +p^{\beta} \sigma ^{\alpha \gamma} r_{\gamma}) \nonumber\\
&+& i
D(t,x_P,l_\perp)\epsilon^{\alpha\beta\delta\rho}p_{\delta}\gamma_{\rho}\gamma_{5}
+....
\end{eqnarray}
Here $m$ is the proton mass. In the matrix structure (\ref{ver})
we wrote only the terms with the maximal powers of a large proton
momentum $p$. The structure functions in (\ref{ver}) are dependent
on the transverse part of the gluon momentum $l_\perp$. The first
two terms of the vertex (\ref{ver}) are symmetric in the gluon
indices $\alpha,\beta$. The structure proportional to $B(t,...)$
determines the spin-non-flip contribution. The term $\propto
K(t,...)$ leads to the transverse spin-flip at the vertex. The
asymmetric structure in (\ref{ver}) is proportional to $D
\gamma_{\rho}\gamma_{5}$ and can be associated with $\Delta G$. It
should give a visible contribution to the  double spin
longitudinal asymmetry $A_{ll}$. We do not consider this structure
here and concentrate on the transverse effects in the proton. In a
QCD--based diquark model of the proton, the first two terms in
(\ref{ver}) were estimated in the proton-proton scattering
amplitude for moderate momentum transfer \cite{gol_kr}. At small
momentum transfer, such a model calculation is not possible, and
we do not know explicitly the functions $B$, $K$,... in
(\ref{ver}). Note that a coupling similar to (\ref{ver}) was found
in high energy quark-quark scattering when  large-distance effects
were considered in the gluon loops \cite{gol93}.

In what follows, we analyse the $\gamma^* g g \to Q \bar Q$
transition amplitude. The typical momentum of quarks there is
proportional to the photon momentum $q$. In the Feynman gauge, we
can decompose the $g_{\mu\nu}$ tensors from $t$- channel gluons
into the longitudinal and transverse parts \cite{grib77}
\begin{equation}\label{gmunu}
g^{\alpha \alpha'}= g^{\alpha \alpha'}_l+g^{\alpha
\alpha'}_\perp\; \mbox{with}\; g^{\alpha \alpha'}_l\sim
\frac{q^\alpha p^{\alpha'}}{(pq)}.
\end{equation}
The product of the $g^{\alpha \alpha'}_l$ tensors by the two-gluon
coupling of the proton can be written in the form
\begin{equation}\label{g_vec}
g^{\alpha' \alpha}_l g^{\beta
'\beta}_lV_{pgg}^{\alpha\beta}(p,t,x_P,l_\perp)\propto
p^{\alpha'}p^{\beta'} \left[\frac{ /\hspace{-1.7mm} q}{(pq)} B(t,x_P,l_\perp)
+\frac{i K(t,x_P,l_\perp)}{2 m (pq)} \sigma ^{\beta \gamma}
q_{\beta} r_{\gamma} \right].
\end{equation}
The structure proportional to $D$ is asymmetric in gluon indices.
It will contribute only in the case when one of the gluon tensors
in (\ref{g_vec}) has a transverse component. It can be seen that
the structure in square brackets in (\ref{g_vec}) is related
directly to the definitions of the skewed gluon distribution (see
e.g. \cite{rad}). So, one can conclude that after integration over
the gluon transverse momentum $l_\perp$, we should have
connections:
\begin{eqnarray}\label{spd}
{\cal F}^g_\zeta (\zeta,t) &\propto& \int d^2l_\perp
B(t,\zeta=x_P,l_\perp) \phi(l_\perp,...) \nonumber\\
{\cal K}^g_\zeta (\zeta,t) &\propto& \int d^2l_\perp
K(t,\zeta=x_P,l_\perp) \phi(l_\perp,...),
\end{eqnarray}
and $B$ and $K$ are nonintegrated gluon distribution functions
which describe spin-average and transverse spin effects in the
proton. The universal function  $\phi$  in (\ref{spd}) will be
found later. In future calculations, we use the
$g^{\alpha\alpha'}$ tensor without its decomposition into
longitudinal and transverse parts.

The hadronic tensor is given by
\begin{equation}
\label{wtenz} W^{\alpha\alpha';\beta\beta'}(s_p)= \sum_{spin \;
s_f} \bar u(p',s_f)
 V_{pgg}^{\alpha\alpha'}(p,t,x_P,l) u(p,s_p) \bar u(p,s_p)
V_{pgg}^{\beta\beta'\,+}(p,t,x_P,l') u(p',s_f)
\end{equation}
and is determined by a trace similar to (\ref{lept}).  The
spin--average and spin--dependent hadron tensors are defined as
\begin{equation}
\label{wpm} W^{\alpha\alpha';\beta\beta'}(\pm)=\frac{1}{2} (
W^{\alpha\alpha';\beta\beta'}(+s_p)
 \pm W^{\alpha\alpha';\beta\beta'}(-s_p)).
\end{equation}
This form is written for an arbitrary spin vector $s_p$ and can be
used as well for transversely or longitudinally polarized target.
In the last case, the contribution of $D$ structure should be
considered. For the leading term of spin- average structure $W(+)$
for the ansatz (\ref{ver}) we find
\begin{equation}
\label{w+f} W^{\alpha\alpha';\beta\beta'}(+) = 16 p^{\alpha}
p^{\alpha'} p^{\beta}  p^{\beta'} ( |B|^2 + \frac{|t|}{m^2}
|K|^2).
\end{equation}
Note that we omit, for simplicity here and in what follows, the
arguments of the $B$ and $K$ functions unless it is necessary.
However, we shall remember that the amplitudes $B$ and $K$ depend
on $l$, otherwise the complex conjugate quantities $B^\star$ and
$K^\star$ are functions of $l'$. The obtained equation for the
spin-average tensor coincides in form with the cross section of
the proton off the spinless particle (a meson e.g.). Really, the
meson--proton helicity-non-flip and helicity-flip amplitudes can
be written in terms of the invariant functions $\tilde B$ and
$\tilde K$ which describe spin-non-flip and spin-flip effects
\begin{equation}
\label{fnf} F_{++}(s,t)= i s [\tilde B(t)] f(t);\;\; F_{+-}(s,t)=
i s \frac{\sqrt{|t|}}{m} \tilde K(t) f(t),
\end{equation}
where $f(t)$ is determined by the Pomeron coupling with meson. The
functions $\tilde B$ and $\tilde K$ are defined by integrals like
(\ref{spd}). The cross-section is written in the form
\begin{equation}
\label{mp_ds} \frac{d\sigma}{dt} \sim  [|\tilde B(t)|^2+
\frac{|t|}{m^2} |\tilde K(t)|^2] f(t)^2.
\end{equation}
The term proportional to $\tilde B$ represents the standard
Pomeron coupling that leads to the non-flip amplitude. The $\tilde
K$ function is the spin--dependent part of the Pomeron coupling
which produces in our case the  spin--flip effects non vanishing
at high-energies. The models \cite{gol_kr,gol_mod} predict a value
of single spin transverse asymmetry of about 10\% for $|t| \sim 3
\mbox{GeV}^2$ which is of the same order of magnitude as that
observed experimentally \cite{krish}. It has been found in
\cite{gol_kr,gol_mod} that the ratio $|\tilde K|/|\tilde B| \sim
0.1$ and has a weak energy dependence. The weak energy dependence
of spin asymmetries in exclusive reactions is not in contradiction
with the experiment \cite{gol_mod,akch}.

The  spin-dependent part of the hadron tensor can be written as
\begin{equation}
\label{w-f1} W^{\alpha\alpha';\beta\beta'}(-) =
S_0^{\alpha\alpha';\beta\beta'} +
S_r^{\alpha\alpha';\beta\beta'}+A_t^{\alpha\alpha';\beta\beta'}.
\end{equation}
The functions $S$ are symmetric in $\alpha,\alpha'$ and $\beta,\beta'$ indices
\begin{equation}
\label{af} S_0^{\alpha\alpha';\beta\beta'}=8 i  \frac{B K^*-B^* K}{m}  p^{\beta}
p^{\beta'} \Gamma^{\alpha \alpha'}
\end{equation}
and
\begin{eqnarray}\label{sr}
S_r^{\alpha\alpha';\beta\beta'}&=&2 i  \frac{B^* K}{m}\left(p^{\alpha}
(r_P)^{\alpha'}+p^{\alpha'} (r_P)^{\alpha} \right)\Gamma^{\beta \beta'}\nonumber\\
&-& 2 i  \frac{B K^*}{m}\left(p^{\beta}
(r_P)^{\beta'}+p^{\beta'} (r_P)^{\beta} \right)\Gamma^{\alpha \alpha'}
\end{eqnarray}
Here
\begin{equation}
\Gamma^{\alpha \alpha'}=p^{\alpha}\epsilon^{\alpha'\gamma\delta\rho} p_{\gamma}
(r_P)_{\delta}
(s_p)_{\rho}+p^{\alpha'}\epsilon^{\alpha\gamma\delta\rho}
p_{\gamma} (r_P)_{\delta}(s_p)_{\rho}
\end{equation}
The function $A_t$ is asymmetric in indices
\begin{eqnarray}
\label{afa} A_t^{\alpha\alpha';\beta\beta'} &=& 2 i |t| \frac{B^* K}{m}
\left[ p^{\alpha} p^{\beta}
\epsilon^{\alpha'\beta'\delta\rho} p_{\delta} (s_p)_{\rho} +
p^{\alpha} p^{\beta'} \epsilon^{\alpha'\beta\delta\rho} p_{\delta}
(s_p)_{\rho}  \right.    \nonumber\\  &+&\left. p^{\alpha'}
p^{\beta} \epsilon^{\alpha\beta'\delta\rho} p_{\delta}
(s_p)_{\rho} + p^{\alpha'} p^{\beta'}
\epsilon^{\alpha\beta\delta\rho} p_{\delta} (s_p)_{\rho} \right]
\end{eqnarray}
Note that these forms are general and can be used for different
polarization vectors of the proton. For longitudinal proton
polarization, the structure $D$ should be considered in addition.

\section{Diffractive Vector Meson Leptoproduction}\label{sect4}

Now we proceed to analyze of the amplitude of  vector meson
production through the photon-two gluon fusion. In what follows,
we will regard mainly the $J/\Psi$ meson production. This meson
can be considered  as an $S$-wave system of heavy $c \bar c$
quarks \cite{berger}. The $J/\Psi$-wave function in this case has
the form
\begin{equation}\label{wf}
\Psi_V=g (/\hspace*{-0.20cm} k+m_q) \gamma_\mu
\end{equation}
 where $k$ is the momentum of a quark, and $m_q$ is its mass. In the
nonrelativistic approximation, both the quarks have the same
momenta $k$ equal to half of the vector meson momentum $K_J$, and
the mass of $c$ quark is equal to $m_J/2$. The transverse quark
motion is not considered. This means that the vector meson
distribution amplitude is approximated by the simple form
$\delta(\tau-1/2)\delta(k_t^2)$. The constant $g$ in the wave
function can be expressed through the $e^+ e^-$ decay width of the
$J/\Psi$ meson \cite{rys}
\begin{equation}
g^2=\frac{3 \Gamma^J_{e^+ e^-} m_J}{64 \pi \alpha^2}.
\end{equation}
The leading twist wave function (\ref{wf}) produces both
amplitudes with longitudinal and transverse vector meson
polarization because of nonzero mass $m_q$. For the light meson
production $m_q=0$,  and one must consider the higher twist
effects to calculate the amplitude with transverse vector meson
polarization (see e.g. \cite{mank}).

\begin{figure} \epsfxsize=9cm \centerline{\epsfbox{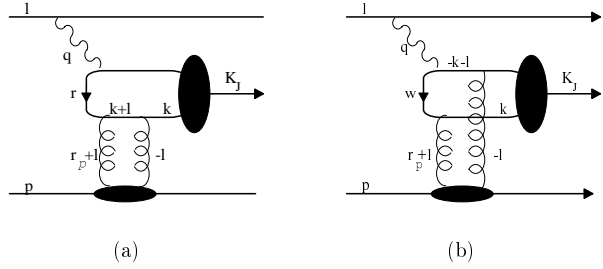}}
\caption{Two-gluon contribution to  diffractive vector meson
production..}
\label{fjpsi}       
\end{figure}

It is known (see \cite{rys,J-Psi} e.g.) that the leading terms of
the amplitude of diffractive vector meson production is mainly
imaginary. We shall consider here only the imaginary parts of the
amplitudes. In this case, only the graphs of Fig.2 contribute. The
gluons are coupled with  single and different quarks in the $c
\bar c$ loop (see Fig. \ref{fjpsi} a, b).
To calculate the imaginary part of the amplitude, we should
consider the $\delta$-function contribution in the $s$-channel
propagators ($k+l$ and $p'-l$ lines for  Fig \ref{fjpsi}\ a). With
the help of  $\delta$ functions the integration over $l$
\begin{equation}
\int d^4 l= \frac{1}{2} \int d l_{+}  d l_{-}  d l_{\perp}
\end{equation}
can be carried out over $l_{+}$ and $l_{-}$ variables. One can
find that both the $l_{\pm}$ components of the vector $l$ are
small: $l_{+} \sim l_{-} \propto 1/p_{+}.$ This results in the
transverse character of the gluon momentum $l^2 \simeq
-l_{\bot}^2$. The same is true for integration over $l$ in the
nonplanar graph of Fig \ref{fjpsi}.\ b. For the arguments in the
off mass shell quark propagators of Fig \ref{fjpsi}.\ a, b, we
find
\begin{eqnarray}
\label{w}
r^2-m_q^2 &=& -\frac{M_J^2+Q^2+|t|}{2}, \nonumber\\
w^2-m_q^2 &=& -2 \left(l_{\bot}^2 + \vec l_{\bot} \vec r_{\bot}
+\frac{M_J^2+Q^2+|t|}{4} \right).
\end{eqnarray}
Thus, these quark lines are far from the mass shell for heavy
vector meson production even for small $Q^2$ \cite{rys}.

In what follows we calculate  the polarized cross section of
vector meson production. The cross section can be represented as a
square of the  $\gamma^\star gg \to V$ amplitude convoluted with
the lepton and hadron polarized tensors. Some details of
calculations conducted for longitudinal target polarization can be
found in \cite{goljpsi}.

We consider both longitudinal and transverse polarization of the
vector meson which can be carried out directly for $J/\Psi$
production for the wave function (\ref{wf}). For the sum over
polarization of $J/\Psi$ polarized vectors $e_J$ we have
\begin{equation}
\label{e_j} \sum_{Spin_J} e_J^{\rho} (e_J^{\sigma})^+=-g^{\rho
\sigma}+ \frac{K_J^{\rho} K_J^{\sigma}}{m_J^2}.
\end{equation}

The spin-average and spin-dependent cross-sections of  vector
meson leptoproduction with longitudinal polarization of a lepton
and transverse polarization of the proton are determined by the
relation
\begin{equation}\label{ds0}
d \sigma(\pm) =\frac{1}{2} \left( d \sigma({\rightarrow}
{\Downarrow) \pm  d \sigma({\rightarrow} {\Uparrow}})\right).
\end{equation}
The cross section $d \sigma(\pm)$ can be written in the form
\begin{equation}\label{ds}
\frac{d\sigma^{\pm}}{dQ^2 dy dt}=\frac{|T^{\pm}|^2}{32 (2\pi)^3
 Q^4 s^2 y}.
\end{equation}
 For the spin-average  amplitude squared we find
\begin{equation}\label{t+}
|T^{+}|^2=  \frac{s^2\, N }{4 \bar Q^4}\,\left( (1+(1-y)^2) m_V^2
+ 2(1 -y) Q^2 \right) \left[ |\tilde B|^2+|\tilde K|^2
\frac{|t|}{m^2} \right].
\end{equation}
Here $\bar Q^2=(m_V^2+Q^2+|t|)/4$, and  $N$ is the normalization
factor
\begin{equation}
\label{n} N=\frac{\Gamma^J_{e^+ e^-} M_J \alpha_s^4 }{27 \pi^2}.
\end{equation}

The term proportional to $(1+(1-y)^2) m_V^2$ in (\ref{t+})
represents the contribution of the virtual photon with transverse
polarization. The $2(1 -y) Q^2$ term describes the effect of
longitudinal photons. This contribution is predominant for high
$Q^2$. The $\tilde B$ and $\tilde K$ functions are expressed
through the integral over the transverse momentum of the gluon.
The function $\tilde B$ is determined by
\begin{eqnarray}\label{bt}
\tilde B&=&\ \bar Q^2 \int \frac{d^2l_\perp (l_\perp^2+\vec
l_\perp \vec r_\perp) B(t,l_\perp^2,x_P,...)}
{(l_\perp^2+\lambda^2)((\vec l_\perp+\vec
\Delta)^2+\lambda^2)[l_\perp^2+\vec l_\perp \vec r_\perp
+\bar Q^2]} \nonumber\\
&\sim& \int^{l_\perp^2<\bar Q^2}_0 \frac{d^2l_\perp
(l_\perp^2+\vec l_\perp \vec r_\perp) }
{(l_\perp^2+\lambda^2)((\vec l_\perp+\vec r_\perp)^2+\lambda^2)}
B(t,l_\perp^2,x_P,...),
\end{eqnarray}
We find that cross sections depend on the variable $\bar Q^2$
which is a modification of the scale variable proposed in
\cite{rys,J-Psi} for the case of large momentum transfer. The term
$(l_\perp^2+\vec l_\perp \vec r_\perp)$ appears in the numerator
of (\ref{bt}) because of the cancellation between the planar and
nonplanar graphs where gluons are coupled with  single and
different quarks (Fig.2). The $\tilde K$ function is determined by
a similar integral. The integral (\ref{bt}) can be connected with
the gluon SPD as
\begin{equation}\label{fspd}
{ {\cal F}^g_{x_P}(x_P,t,\bar Q^2)} \sim \int^{l_\perp^2<\bar
Q^2}_0 \frac{d^2l_\perp (l_\perp^2+\vec l_\perp \vec r_\perp) }
{(l_\perp^2+\lambda^2)((\vec l_\perp+\vec r_\perp)^2+\lambda^2)}
B(t,l_\perp^2,x_P,...) = \tilde B.
\end{equation}
Note that if one considers effects of the transverse quark motion
in the the vector meson wave function, the scale variable in SPD
will be changed to $\bar Q^2 \to \bar Q^2+k_\perp^2$
\cite{rys1,golwf}. Thus, the $B(l_\perp^2,x_P,...)$ function is a
nonintegrated spin- average gluon distribution. The $\tilde K$
function is proportional to the ${\cal K}^g_{x_P}(x_P,t)$
distribution. The function $\phi$ in (\ref{spd})  has the form
\begin{equation}\label{fi}
\phi(l_\perp,...)=\frac{(l_\perp^2+\vec l_\perp \vec r_\perp) }
{(l_\perp^2+\lambda^2)((\vec l_\perp+\vec r_\perp)^2+\lambda^2)}.
\end{equation}

The spin-dependent amplitude squared looks like
\begin{equation}\label{t-}
|T^{-}|^2=  \frac{\vec Q \vec S_\perp}{4 m}\; \frac{s |t| N}{4
\bar Q^4}\left(Q^2+ m_V^2+|t| \right) \frac{\tilde B \tilde K^*+
\tilde B^* \tilde K }{2} .
\end{equation}
We shall use the spin-dependent cross sections obtained here for
the numerical analysis of polarized vector meson production in
section 6.
\section{Diffractive $Q \bar Q$ Photoproduction}

Let us study now the diffractive $Q \bar Q$ production in the
lepton-proton reaction. This process  is determined by similar
graphs shown in Fig. \ref{fjpsi}. The change is in the
photon-two-gluon fusion amplitude where we do not project the $Q
\bar Q$ state onto the vector meson. The quark-antiquark
contribution, instead of $t$- channel gluons, is possible for
light quark production. To suppress this contribution of the quark
structure function which should be essential at large $x$, we
investigate quark production at small $x\le 0.1$. In this
kinematical region the gluon contribution is predominant.
\begin{figure}[h]
\epsfxsize=7cm \centerline{\epsfbox{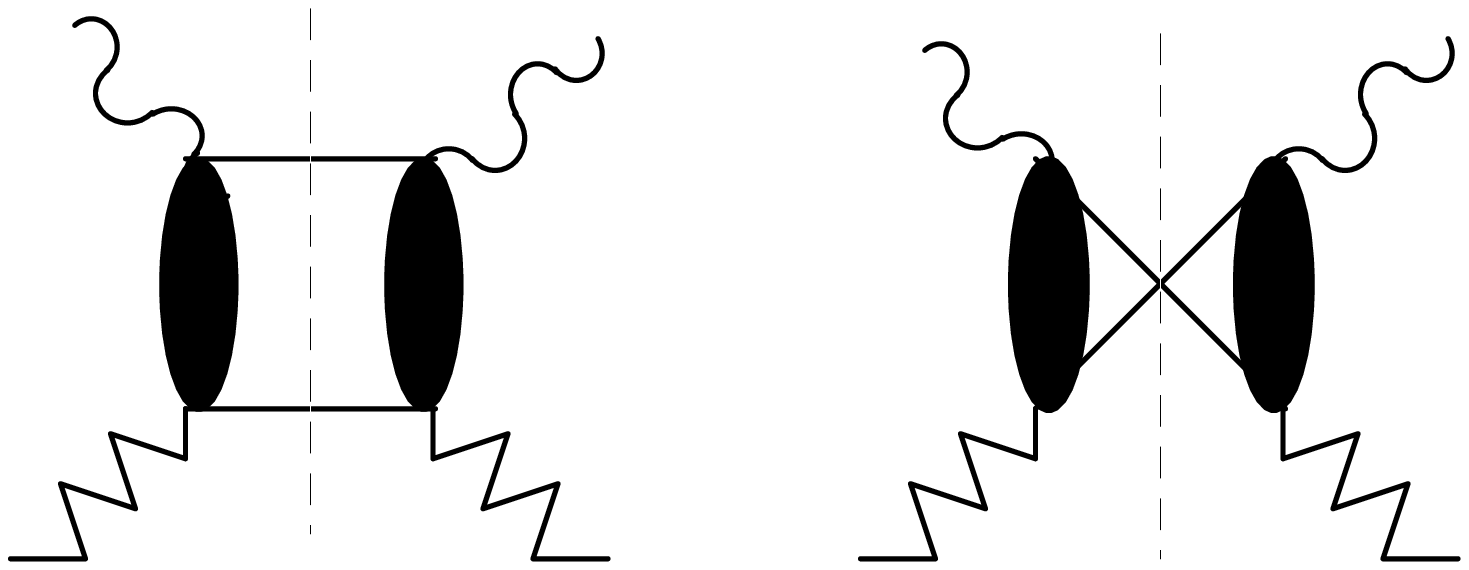}} \caption{Box
graphs contributions to the cross section of diffractive $Q \bar
Q$ production.}
\label{fbox}       
\end{figure}

\begin{figure}[h]
\epsfxsize=7cm \centerline{\epsfbox{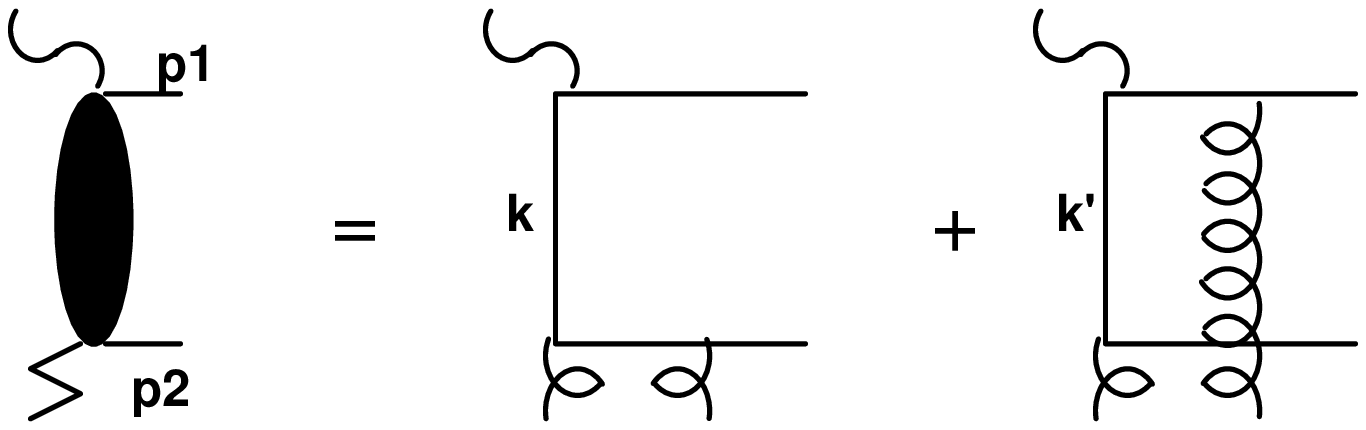}} \caption{The
amplitude of photon-two-gluon fusion}
\label{fgqq}       
\end{figure}

As in the case of vector meson production, we  calculate the spin
average and spin-dependent cross section (\ref{ds}) of diffractive
$Q \bar Q$ leptoproduction. To calculate these cross sections, we
should integrate the corresponding amplitudes squared over the $Q
\bar Q$ phase space $dN_{Q \bar Q}=\Pi_f\frac{d^3p_f}{(2 \pi)^3 2
E_f}$ with the delta function that reflects the momentum
conservation. It can be easily seen that
\begin{equation}\label{cross}
\frac{d^3p_1}{2 E_1} \frac{d^3p_2}{2 E_2} \delta^4(q+r_P-p_1-p_2)=
d^4p_1 \delta(p_1^2-m_q^2) \delta(p_2^2-m_q^2),
\end{equation}
and the calculation of $\gamma gg \to Q \bar Q$ cross section is
equivalent to computation of the imaginary part of the quark loop
diagram shown in Fig.\ref{fbox}. The amplitude of photon-two-gluon
fusion shown by  blobs in Fig.\ref{fbox} represents a sum of
graphs in Fig.\ref{fgqq}. The diagrams of Fig.\ref{fgqq} are
similar to the planar and nonplanar gluon graphs of
Fig.\ref{fjpsi}. As a result, the gluon contribution to the cross
section should be similar to that obtained in (\ref{bt}).
\vspace{5mm}

The final quark momenta $p_1$, $p_2$ and the momentum of the
off-mass-shell quark $k$ $(k')$ can be determined with the help of
delta functions in (\ref{cross}). There are two integration
regions for $k$ vectors. For  Region I, we find that $p_1 \sim q$
and $p_2 \sim r_P$
\begin{eqnarray}\label{p1p2}
p_1& \sim &\left(y p_{+} -\frac{|t|}{p_{+}}-
\frac{m_q^2+(\vec r_\perp+\vec k_\perp)^2}{p_{+} x_P},
\frac{m_q^2+(\vec q_\perp-\vec k_\perp)^2}{p_{+} y},
(\vec q_\perp-\vec k_\perp)  \right),
\nonumber \\
p_2& \sim &\left(\frac{m_q^2+(\vec r_\perp+\vec k_\perp)^2}{p_{+}
x_P}, x_P p_{+}-\frac{Q^2}{p_{+}}-\frac{m_q^2+(\vec q_\perp-\vec
k_\perp)^2}{p_{+} y}, (\vec r_\perp+\vec k_\perp) \right),
\end{eqnarray}
and vector $k$ is mainly transverse: $k^2 \sim
-k_\perp^2$. The $t$ channel gluon contribution is predominated if
both quark-proton energies are large. We find that
\begin{equation}\label{energy}
(p_1+p)^2 \sim y p_{+}^2 \;\;\; (p_2+p)^2 \sim
\frac{m_q^2+k_\perp^2}{x_P}.
\end{equation}
Thus, $x_P$ should be quite  small ($x_P \le 0.1$). At the same,
time $k_\perp^2$ should be not small (we shall use $k_\perp^2 > 1
\mbox{GeV}^2$). A non small value for $k^2$ produces the large
quark virtuality for the graphs of Fig. (\ref{fgqq}).

For Region II, the quarks momentum changed places $p_1
\leftrightarrow p_2$. In this case, the vector $k$ has a large
longitudinal component, and $k^2 \sim -x_P p_+^2$. One can suppose
that such contributions should be suppressed. However, in this
case, we find a similar large variable from trace in the numerator
of the diagram that compensates $p_+^2$ in denominator.  A similar
compensation between the numerator and denominator takes place for
the nonplanar quark loop diagrams (the second graph in
Fig.\ref{fbox}). In this case, we have a large variable only in
one propagator. The calculation of  $\gamma gg \to Q \bar Q$
process is more complicated than the  vector meson case. We must
consider here 8 graphs with two Regions for quark momenta. This
generates a complete set of graphs of $Q \bar Q$ production.

The integration over quark momenta $k_\pm$ in the loop can be
carried out with the help of delta functions in (\ref{cross})
\begin{equation}\label{int}
d^4k \;\delta(p_1^2-m_q^2)\;\delta(p_2^2-m_q^2) \sim
\frac{d^2k_\perp}{M_X^2 \sqrt{1-4 (k_\perp^2+m_q^2)/M_X^2}}
\end{equation}

As a result, the spin-average and spin-dependent cross section can
be written in the form
\begin{equation}
\label{sigma} \frac{d^5 \sigma(\pm)}{dQ^2 dy dx_p dt dk_\perp^2}=
\left(^{(2-2 y+y^2)} _{\hspace{3mm}(2-y)}\right)
 \frac{C(x_P,Q^2) \; N(\pm)}
{\sqrt{1-4(k_\perp^2+m_q^2)/M_X^2}}.
\end{equation}
Here $C(x_P,Q^2)$ is a normalization function which is common for
the spin average and spin dependent cross section; $N(\pm)$ is
determined by a sum of graphs in Fig. \ref{fbox}, \ref{fgqq}
integrated over the gluon momenta $l$ and $l'$
\begin{eqnarray}
\label{tpm} N(\pm) = \int \frac{ d^2 l_{\bot} d^2 l_{\bot}'
(l_\perp^2+\vec l_\perp \vec r_\perp)\; ((l'_\perp)^2+\vec
l'_\perp \vec r_\perp)\;D^{\pm}(t,Q^2,l_{\bot},l'_{\bot},\cdots)}
{(l_\perp^2+\lambda^2)((\vec l_\perp+\vec r_{\perp})^2+\lambda^2)
(l_\perp^{'2}+\lambda^2)((\vec l'_\perp+\vec
r_{\perp})^2+\lambda^2)}.
\end{eqnarray}
The $D^{\pm}$ function here is a sum of traces over the quark
loops of the graphs in Figs. \ref{fbox}, \ref{fgqq} convoluted
with the spin average and spin-dependent tensors. The calculation
shows a considerable cancellation between the  planar and
nonplanar contribution of the graphs in Fig.\ref{fgqq}. As a
result, in the numerator of (\ref{tpm}) we find the terms
proportional to the gluon momenta $l_{\perp}$ and $l'_{\perp}$ as
in the case of vector meson production (\ref{bt}).

For simplicity we write the analytic forms of the graph
contribution to the cross sections in the limit $\beta \to 0$. The
numerical calculation will be fulfiled for arbitrary $\beta$. The
contribution of the sum of the graphs of Figs. \ref{fbox},
\ref{fgqq} to the $D^{+}$ function for Region I can be written in
the form
\begin{equation}\label{d+}
D_I^{+}=\frac{Q^2 \left(|B|^2+|t|/m^2 |K|^2
\right)\left((k_\perp+r_\perp)^2+m_q^2 \right)}
{\left(k_\perp^2+m_q^2 \right)
\left((k_\perp-l_\perp)^2+m_q^2\right)
\left((k_\perp-l'_\perp)^2+m_q^2\right)}.
\end{equation}
This function contains a product of the off-mass-spell quark
propagators in the graphs Fig. \ref{fbox}, \ref{fgqq}. We see that
the quark virtuality here is quite different as compared to the
vector meson case. We have no the terms proportional to $Q^2$ as
in (\ref{w}).

This will change the scale in corresponding gluon structure
functions. Really, denominators in (\ref{d+}) determine the
effective integration region over $l$ and $l'$ in (\ref{n}). We
can rewrite approximately the contribution of $D^p(+)$ to $N(+)$
\begin{equation}\label{nn}
N^p(+) \sim \frac{\left(|\tilde B|^2+|t|/m^2 |\tilde K|^2 \right)
\left((k_\perp+r_\perp)^2+m_q^2 \right)}{\left(k_\perp^2+m_q^2
\right)^3}
\end{equation}
with
\begin{equation}\label{bqq}
\tilde B \sim \int^{l_\perp^2<k_0^2}_0 \frac{d^2l_\perp
(l_\perp^2+\vec l_\perp \vec r_\perp) }
{(l_\perp^2+\lambda^2)((\vec l_\perp+\vec r_\perp)^2+\lambda^2)}
B(t,l_\perp^2,x_P,...) =  {\cal F}^g_{x_P}(x_P,t,k_0^2).
\end{equation}
and
\begin{equation}\label{kqq}
\tilde K \sim \int^{l_\perp^2<k_0^2}_0 \frac{d^2l_\perp
(l_\perp^2+\vec l_\perp \vec r_\perp) }
{(l_\perp^2+\lambda^2)((\vec l_\perp+\vec r_\perp)^2+\lambda^2)}
K(t,l_\perp^2,x_P,...) =  {\cal K}^g_{x_P}(x_P,t,k_0^2)
\end{equation}
with $k_0^2 \sim k_\perp^2+m_q^2$. For nonzero $\beta$ this scale
is changed to $k_0^2 \sim \frac{k_\perp^2+m_q^2}{1-\beta}$ and
coincides with that found in \cite{bart96, ryskin97}. As we have
expected, the gluon structure functions are determined by the same
integrals as in (\ref{fspd}) but on a different scale.
\bigskip
\begin{figure}[h]
\epsfxsize=10,8cm \centerline{\epsfbox{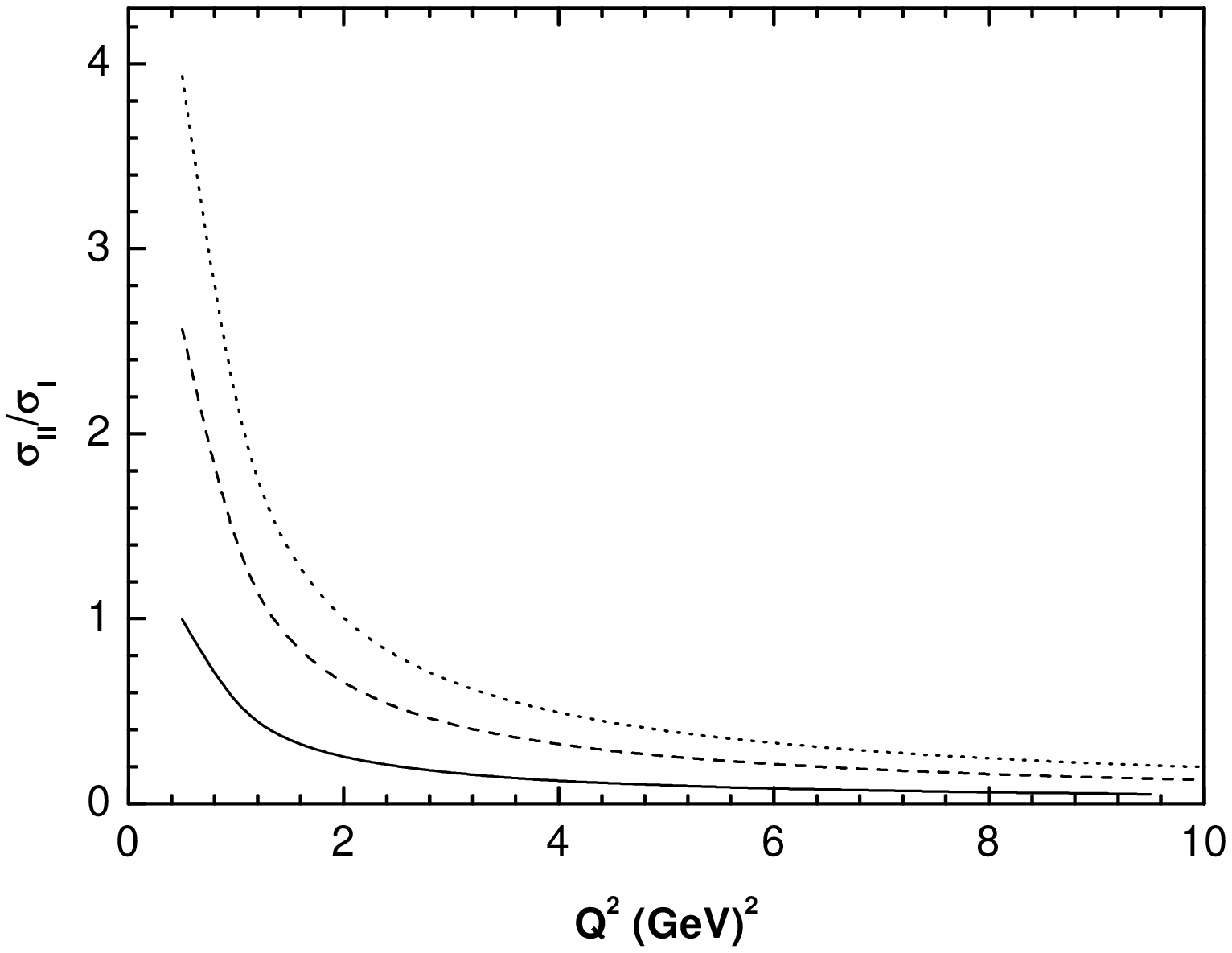}} \caption{The
ratio of cross sections for Regions II and I at $\sqrt{s}=20
\mbox{GeV}$ for $x_P=0.1$, $y=0.5$, $|t|=0.3\mbox{GeV}^2$: solid
line- for $k^2=1 \mbox{GeV}^2$, dashed line- for $k^2=3
\mbox{GeV}^2$, dotted line- for $k^2=5 \mbox{GeV}^2$.}
\label{sir}       
\end{figure}

For Region II, only the first planar graph of  Fig. \ref{fbox}
contributes. As a result of compensation of the gluon
contributions shown in Fig. \ref{fgqq}, the form (\ref{tpm}) is
valid here too. The graphs here have lines with large quark
virtuality. Propagators of these lines become  pointlike. As a
result, the contribution to the cross section for Region II has
different $Q^2$ and $k^2$ dependence with respect to Region I. We
find
\begin{equation}\label{d2}
D_{II}^{+}= \frac{2 (1-y) \left(|B|^2+|t|/m^2 |K|^2 \right) }{(2-2
y+y^2) \left((k_\perp+r_\perp)^2+m_q^2 \right)}.
\end{equation}
The ratio of the cross sections for Regions I and II is shown in
Fig \ref{sir}. The ratio is growing with $k^2$. We find that the
integration region II is essential at small $Q^2$. It can be seen
that the contribution of the $D$ functions (\ref{d+},\ref{d2}) to
the cross section (\ref{sigma}) is proportional to $D_{I}^{+}
\propto (2-2 y+y^2)$, $D_{II}^{+} \propto 2 (1-y)$. They
represent, as in (\ref{t+}), the contributions with transverse and
longitudinal photon polarization, respectively \cite{bart96}.

The contribution of all graphs to the function $N(+)$ can be written as
\begin{equation}\label{np}
N(+)=\left(|\tilde B|^2+|t|/m^2 |\tilde K|^2 \right)
\Pi^{(+)}(t,k_\perp^2,Q^2)
\end{equation}
The function $\Pi^{(+)}$ for nonzero $\beta$
is complicated in form and will be calculated numerically.

The same procedure is used in calculations of the spin-dependent
cross sections. The spin-dependent leptonic and hadronic tensors
(\ref{lpm},\ref{w-f1}) are used in this case. The spin-dependent
part of the hadronic tensor (\ref{w-f1}) is much more complicated
as compared with (\ref{w+f}). This results in tangled analytic
expressions of the spin-dependent cross sections. We shall discuss
here only the general structure of this observable. In addition to
the term observed in (\ref{t-}) and proportional to the scalar
production $\vec Q \vec S_\perp$, the new term $\propto \vec
k_\perp \vec S_\perp$ appears. As a result, we find the following
representation of the function $N(-)$
\begin{equation}\label{nm}
N(-)=\sqrt{\frac{|t|}{m^2}} \left(\tilde B \tilde K^*+\tilde B^*
\tilde K\right) \left[ \frac{(\vec Q \vec S_\perp)}{m}
\Pi^{(-)}_Q(t,k_\perp^2,Q^2) +\frac{(\vec k_\perp \vec
S_\perp)}{m} \Pi^{(-)}_k(t,k_\perp^2,Q^2)\right].
\end{equation}
The second term in (\ref{nm}) cannot be found in the vector meson
production, because we should integrate there the amplitudes over
$d^2 k_\perp$. The functions $\Pi^{(-)}_{Q(k)}$ will be calculated
numerically as the function $\Pi^{(+)}$.

\section{Numerical Results for Vector Meson Leptoproduction}
\label{sect6}
We shall calculate the polarized cross section (\ref{ds0}) of
diffractive $J/\Psi$ production determined by the amplitudes
(\ref{t+}, \ref{t-}). The spin-average cross section of the vector
meson production at a small momentum transfer is  proportional to
the $|\tilde B|^2$ function (\ref{t+}) which is connected with the
skewed gluon distribution (\ref{fspd}). This result is in
accordance with the imaginary part of the amplitude found on the
basis of the SPD approach \cite{rad}. We use here a simple
parameterization of the SPD as a product of the form factor and
the ordinary gluon distribution
\begin{equation}
\label{b_g}
\tilde B(t,x_P, \bar Q^2) =F_B(t) \left( x_P G(x_P,\bar Q^2)  \right).
\end{equation}
where for simplicity  the form factor $F_B(t)$ is chosen as the
electromagnetic form factor of the proton. Such a simple choice
can be justified by that the Pomeron--proton vertex might be
similar to the photon--proton coupling \cite{lansh-m,nach}
\begin{equation}
\label{fp}
F_B(t) \sim F^{em}_p(t)=\frac{(4 m_p^2+2.8 |t|)}{(4 m_p^2+|t|)(
1+|t|/0.7GeV^2)^2}.
\end{equation}

 To perform simple estimations, we shall use our results from
\cite{gol_kr,gol_mod} where it was found  that the ratio of spin
asymmetries in exclusive reactions at a small momentum transfer
may have a weak energy dependence. The corresponding asymmetries
are proportional to the ratio $|\tilde K|/|\tilde B|$ at small $x
\sim 1/s$. We shall suppose that this is true for the ratio of
spin--dependent and spin--average densities in our case too and
\begin{equation}\label{ratio}
\frac{|\tilde K|}{|\tilde B|} \sim 0.1.
\end{equation}
We shall use this value in our estimation of spin asymmetries of
hadron leptoproduction at small $x$.

The energy dependence of the cross sections is determined by the
Pomeron contribution to the gluon distribution function at small
$x$
\begin{equation}
\label{g_x}
\left(x_P G(x_P,\bar Q^2) \right) \sim \frac{const}{x_P^{1-\alpha_p(t)}}
\sim \left( \frac{s y}{m_J^2+Q^2+|t|} \right) ^{(\alpha_p(t)-1)}.
\end{equation}
Here $\alpha_p(t)$ is a Pomeron trajectory which is chosen in the form
\begin{equation}\label{pom}
  \alpha_p(t)=1+\epsilon+ \alpha' t.
\end{equation}
with $\epsilon=0.15$ and $ \alpha'=0$. These values are in
accordance with the fit of  diffractive $J/\Psi$ production by
ZEUS \cite{zeus_p}

\begin{figure}
\epsfxsize=10cm \centerline{\epsfbox{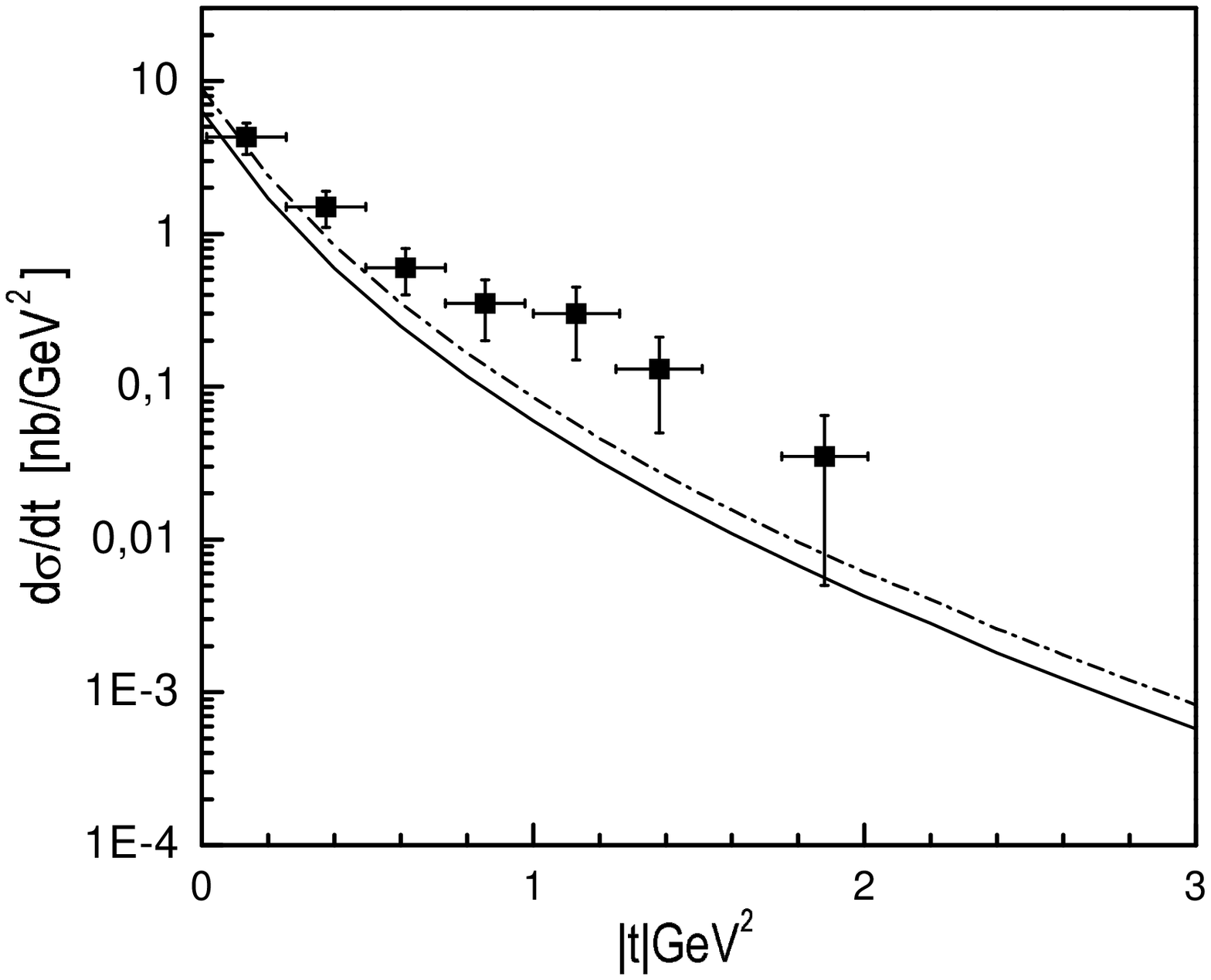}} \caption{The
differential cross section of  $J/\Psi$ production at HERA energy:
solid line -for $|\tilde K|/|\tilde B|=0$; dot-dashed line -for
$|\tilde K|/|\tilde B|=0.1$. Data are from \cite{jpsi1}.}
\label{ds-hr}       
\end{figure}

 The typical scale of the reaction is determined by $\bar
Q^2=(m_J^2+Q^2+|t|)/4$. For not large $Q^2$ and $|t|$,  the value
of $\bar Q^2$ is about 2.5-3.0 $\mbox{GeV}^2$. In this region, we
can work with fixed  $\alpha_s \sim 0.3$. An effective gluon mass
in (\ref{bt}) is chosen to be equal to 0.3 $\mbox{GeV}^2$. The
cross section depends on this parameter weakly. The value of
$\Gamma^J_{e^+ e^-}=5.26 \mbox{keV}$ is used. The predicted cross
sections are shown in Fig. \ref{ds-hr}. Our results reproduce
experimental data quite well.

The $A_{lT}$ asymmetry for vector meson production  is determined
by the ratio of cross sections determined in (\ref{t-},\ref{t+})
\begin{equation}
\label{asylt} A_{lT}=\frac{\sigma(-)}{\sigma(+)} \sim \frac{\vec Q
\vec S_\perp}{4 m}\; \frac{y x_P |t|}{ (1+(1-y)^2) m_V^2 + 2(1 -y)
Q^2} \; \frac{\tilde B \tilde K}{ |\tilde B|^2+|\tilde K|^2
|t|/m^2}.
\end{equation}
For a small momentum transfer,  this asymmetry can be approximated
as
\begin{equation}\label{alt}
 A_{LT} \sim C_g \frac{{\cal K}^g_\zeta(\zeta)}{{\cal F}^g_\zeta(\zeta)}
 \;\;\;\mbox{with} \;\zeta=x_P
\end{equation}
Simple estimations show that the coefficient $C_g(J/\Psi)$ at
HERMES energy for $y=0.5, |t|=1\mbox{GeV}^2, Q^2=5\mbox{GeV}^2$ is
quite small, about 0.007.  To get expected values for coefficients
(\ref{alt}) for light vector mesons, we use the same Eq.
(\ref{asylt}).  The  simple model used for the wave function
predicts weak mass dependence of the gluon contribution to the
asymmetry. For the same kinematical variables, $C(\phi) \sim
C(\rho) \sim 0.008$. However, these results are obtained for the
nonrelativistic meson wave function of the form
$\delta(\tau-1/2)\delta(k_t^2)$ that is not a good approximation
for light meson production. To get suitable predictions for $\rho,
\phi$ meson production, it is important to study a more realistic
wave function and take into consideration the transverse quark
degrees of freedom. Moreover, for $\rho,\phi$ meson production,
the contribution of the quark SPD should be considered in the
HERMES energy range.

\begin{figure}
\epsfxsize=10.8cm \centerline{\epsfbox{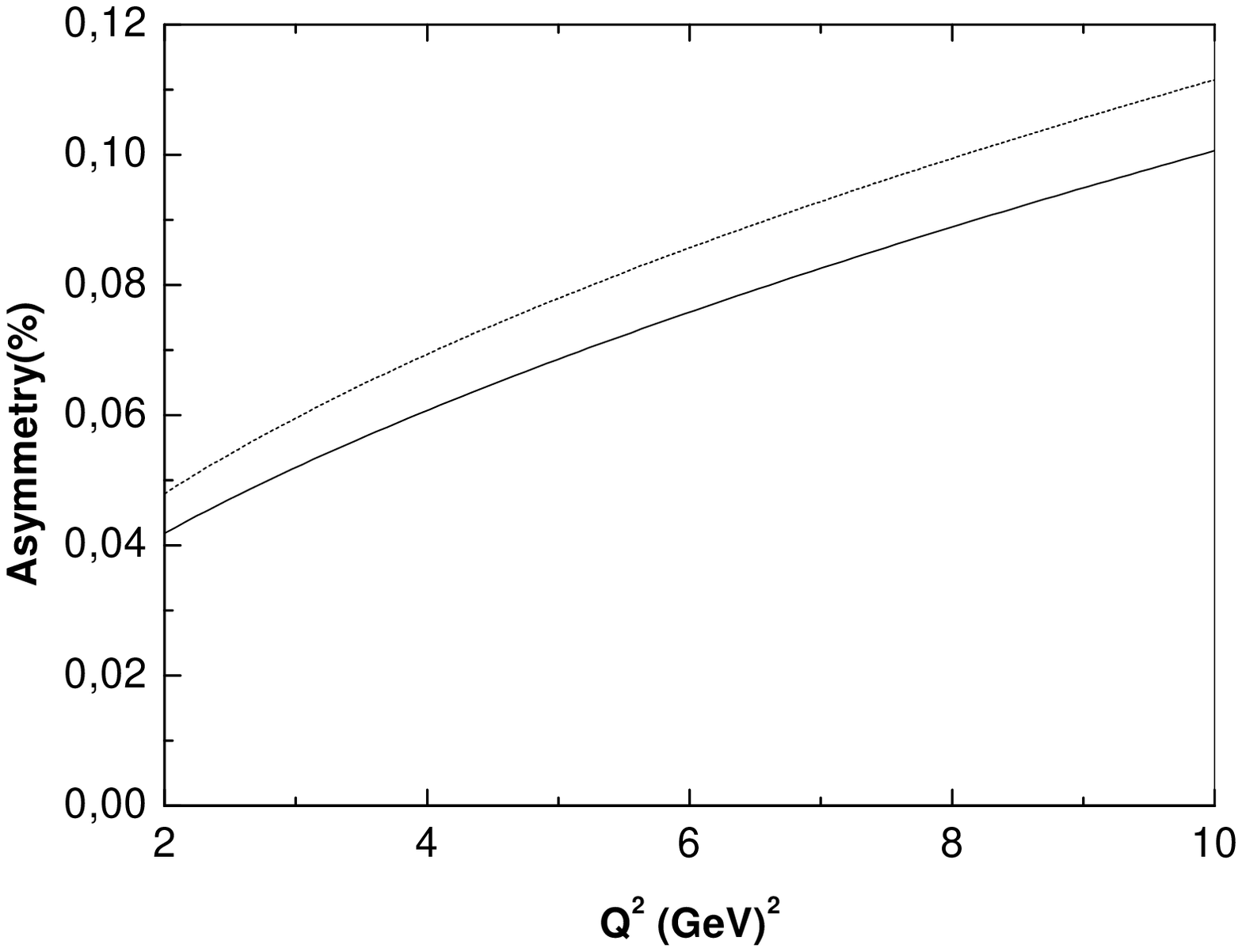}} \caption{The
$A_{lT}$ asymmetry for vector meson production at $\sqrt{s}=7 \mbox{GeV}$
($y$=0.5, $|t|=1 \mbox{GeV}^2$): solid line -for $J/\Psi$ production; dotted
line -for $\rho$ production.}
\label{altj}       
\end{figure}

The asymmetry predicted for $J/\Psi$ production  at HERMES
energies is shown in Fig. \ref{altj}  ($\tilde K/\tilde B=0.1$)
for the case when the transverse part of the photon momentum is
parallel to the target polarization $S_\perp$. Simple estimations
on the basis of (\ref{alt}) for $\rho$ meson production are shown
there too.
 At the HERA energies, asymmetry will be extremely small.

\section{Predictions for  $Q \bar Q$ Leptoproduction}\label{sect7}

 We shall discuss here our prediction for polarized diffractive $Q
\bar Q$ production. We do not consider the cross section but only
the asymmetry $A_{lT}=\sigma(-)/\sigma(+)$.  In estimations we
shall use the same parameterizations of SPD as in (\ref{b_g}) with
the functions determined in (\ref{fp}). As in the case of vector
meson production, the asymmetry is approximately proportional to
the ratio of polarized and spin--average gluon distribution
functions
\begin{equation}\label{cltqq}
 A_{LT}^{Q \bar Q} \sim C^{Q \bar Q} \frac{{\cal K}^g_\zeta(\zeta)}
 {{\cal F}^g_\zeta(\zeta)}
 \;\;\;\mbox{with} \;\zeta=x_P
\end{equation}
As previously, in our estimations we use the value $|\tilde
K|/|\tilde B| \sim 0.1$.

\begin{figure}
\epsfxsize=10,8cm \centerline{\epsfbox{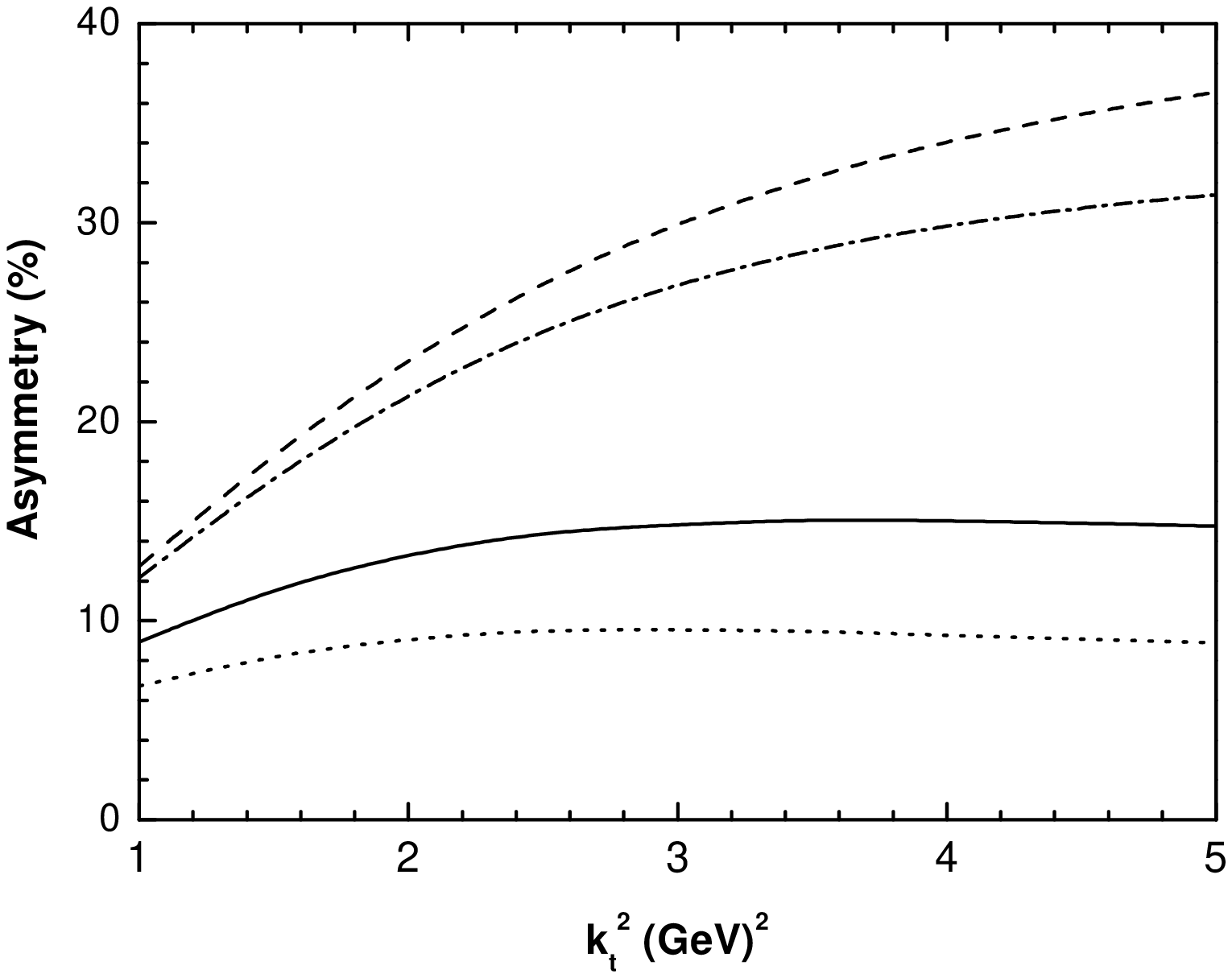}} \caption{The
$A^k_{lT}$ asymmetry in diffractive light $Q \bar Q$ production at
$\sqrt{s}=20 \mbox{GeV}$ for $x_P=0.1$, $y=0.5$, $|t|=0.3 \mbox{GeV}^2$:
dotted line-for $Q^2=0.5 \mbox{GeV}^2$; solid line-for $Q^2=0.5 \mbox{GeV}^2$;
dot-dashed line-for $Q^2=5 \mbox{GeV}^2$; dashed line-for $Q^2=10 \mbox{GeV}^2$.}
\label{altl}       
\end{figure}
\begin{figure}
\epsfxsize=10,8cm \centerline{\epsfbox{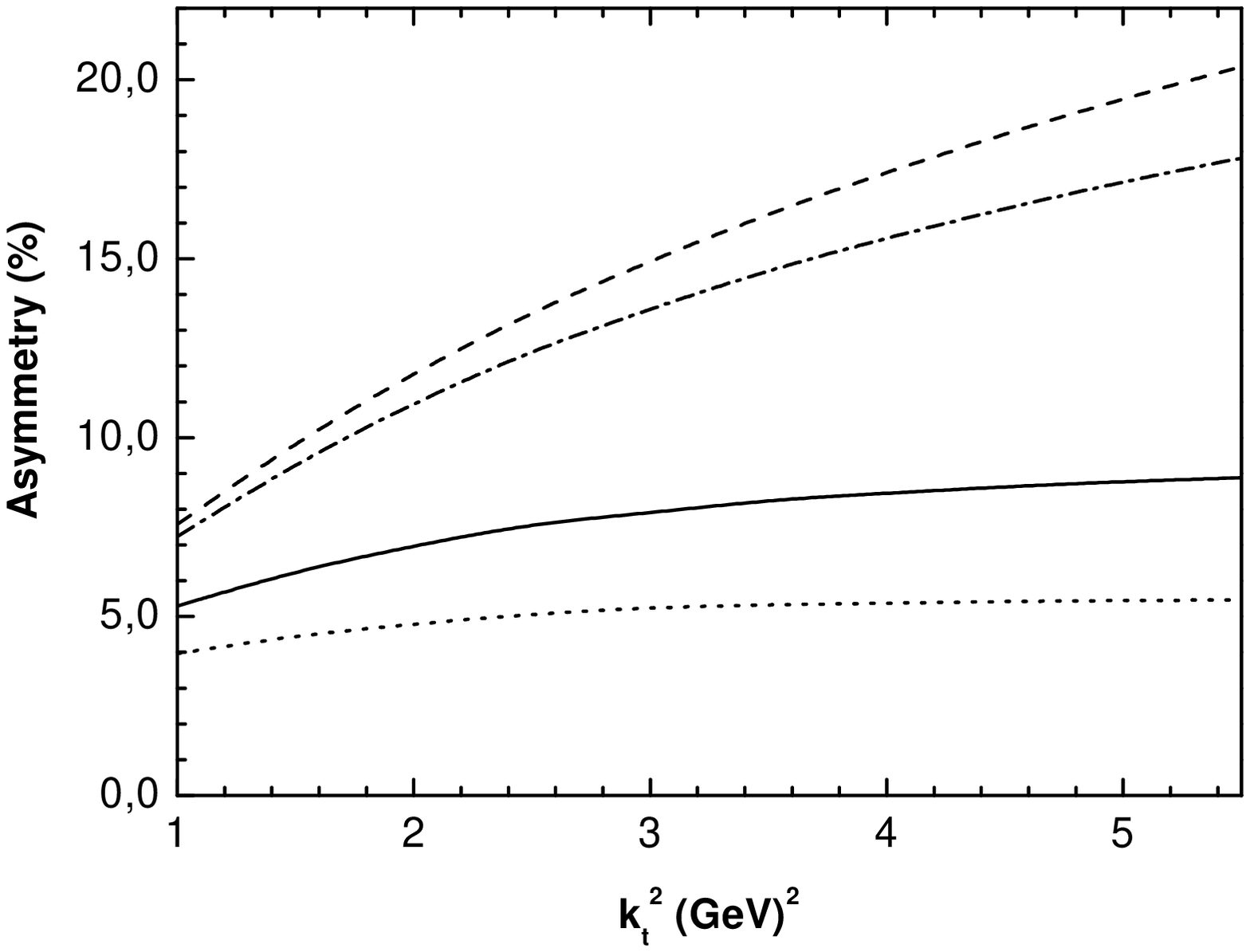}} \caption{The
$A^k_{lT}$ asymmetry in diffractive heavy $Q \bar Q$ production at
$\sqrt{s}=20 \mbox{GeV}$ for $x_P=0.1$, $y=0.5$, $|t|=0.3 \mbox{GeV}^2$:
dotted line-for $Q^2=0.5 \mbox{GeV}^2$; solid line-for $Q^2=0.5 \mbox{GeV}^2$;
dot-dashed line-for $Q^2=5 \mbox{GeV}^2$; dashed line-for $Q^2=10 \mbox{GeV}^2$.}
\label{alth}       
\end{figure}
\begin{figure}
\epsfxsize=10,8cm \centerline{\epsfbox{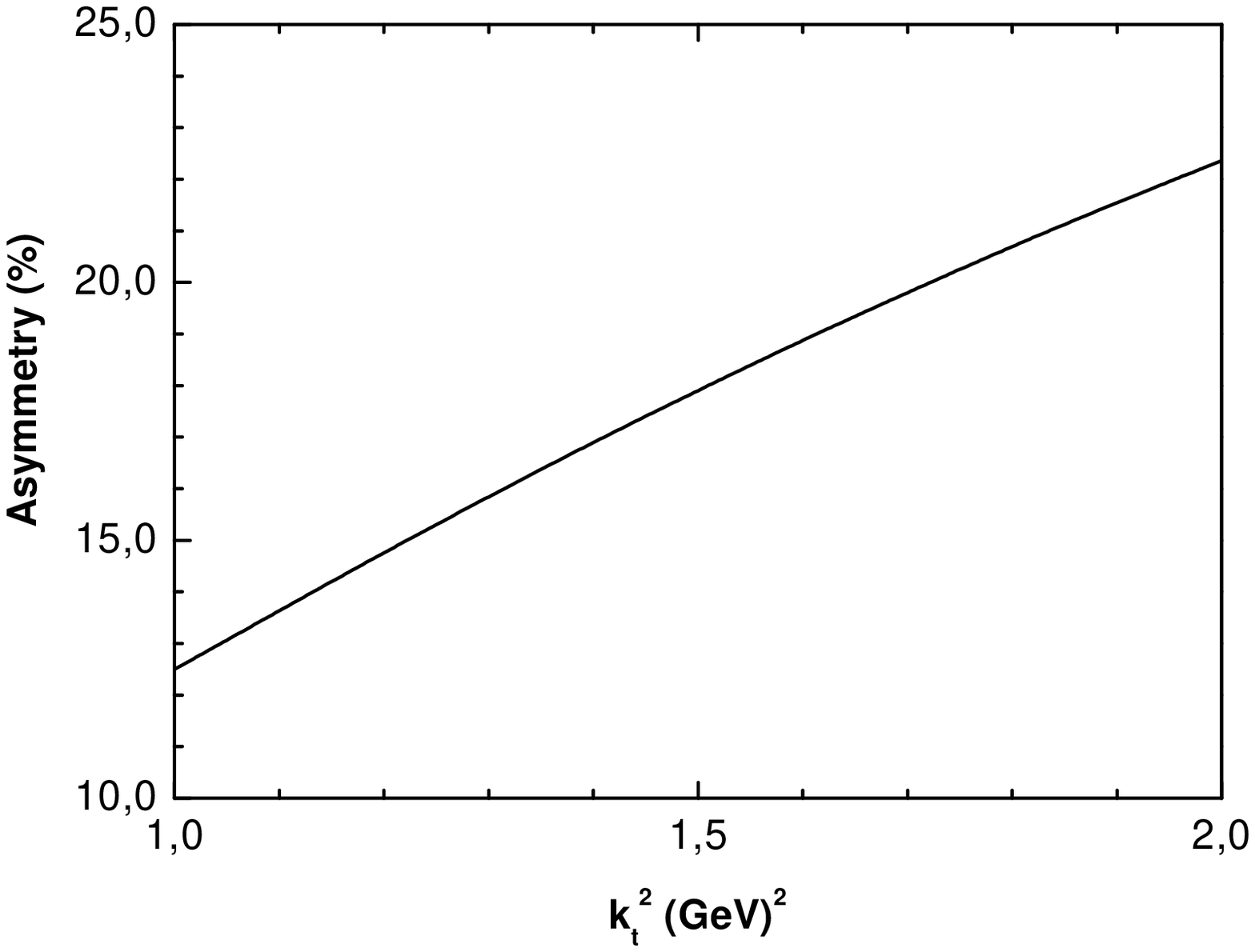}} \caption{The
$A^k_{lT}$ asymmetry  in diffractive light $Q \bar Q$ production
for $Q^2=5\mbox{GeV}^2$, $x_P=0.1$, $y=0.5$,
$|t|=0.3(GeV)^2$ at $\sqrt{s}=7 \mbox{GeV}$.}
\label{althm}       
\end{figure}

The spin-dependent contribution has two terms proportional to the
scalar products $\vec k_\perp \vec S_\perp$ and $\vec Q \vec
S_\perp$ (\ref{nm}). We shall study these contributions to
asymmetry separately. The first term will be analyzed for the case
when the transverse jet momentum $\vec k_\perp$ is parallel to the
target polarization $\vec S_\perp$. The asymmetry is maximal in
this case. We would like to emphasize here that to observe this
contribution to asymmetry, it is necessary to distinguish
experimentally the quark and antiquark jets. This can be realized
presumably by the charge  of the leading particles in the jet
which should be connected in charge with the quark produced in
photon-gluon fusion. This is an indispensable condition in the
experimental study of that asymmetry caused by the fact that the
transverse momentum of a quark and an antiquark produced in the
process are opposite in sign. If we do not separate events with
$\vec k_\perp$ for the quark jet e.g., the resulting asymmetry
will be zero.

The spin--dependent cross section  vanishes for $Q^2 \to 0$, while
the spin--average cross section is constant in this limit. As a
result, the asymmetry can be estimated as $A_{lT} \propto
Q^2/(Q^2+Q^2_0)$ with $Q^2_0 \sim 1 \mbox{GeV}^2$. The $Q^2$
dependence of the asymmetry for light quark production at energy
$\sqrt{s}=20 \mbox{GeV}$ is shown in Fig. \ref{altl}. The
predicted asymmetry for heavy $c \bar c$ production is
approximately of the same order of magnitude (Fig.\ref{alth}).

\begin{figure}
\epsfxsize=10,8cm \centerline{\epsfbox{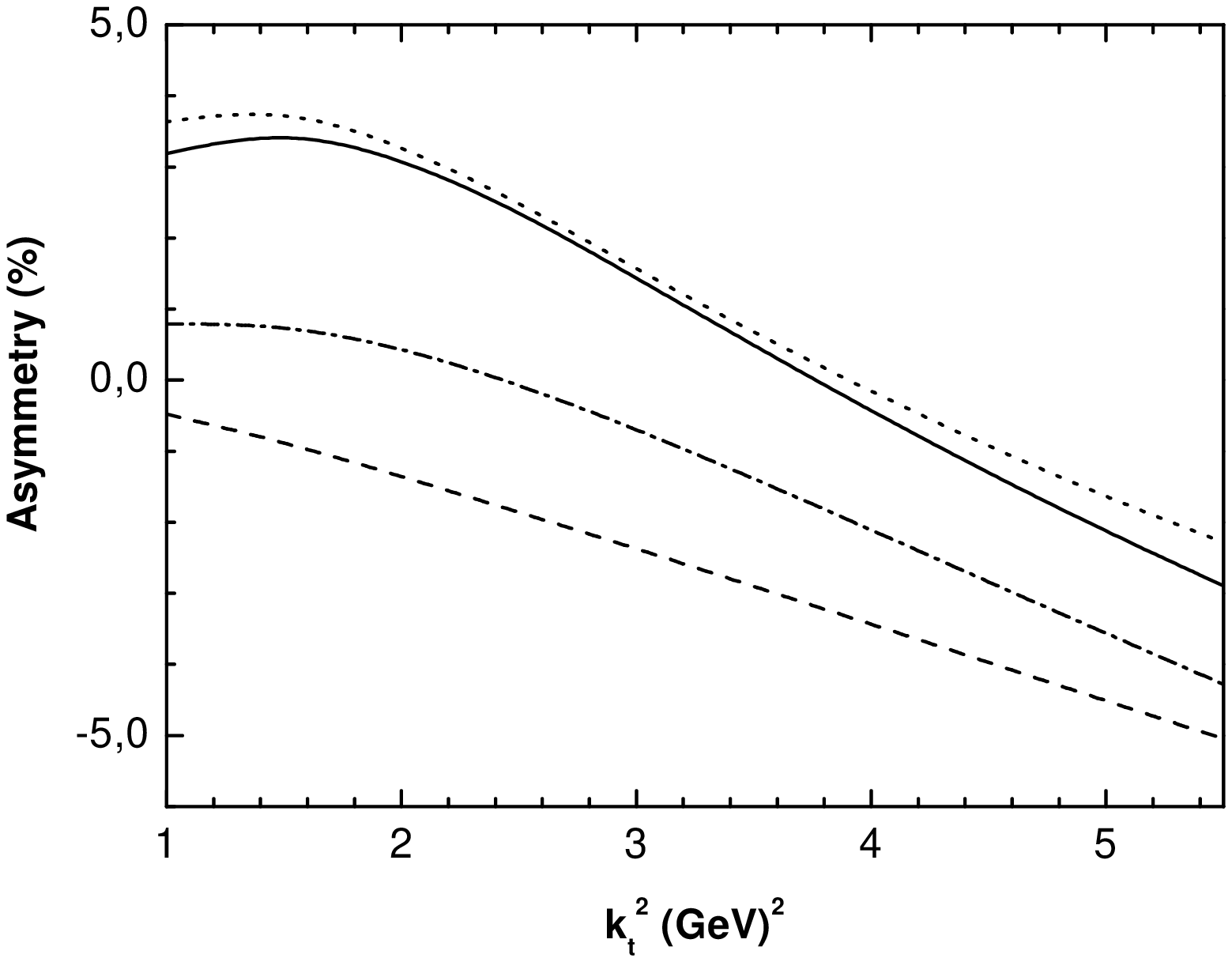}}
\caption{$A^Q_{lT}$ asymmetry in diffractive light $Q \bar Q$ production at
$\sqrt{s}=20 \mbox{GeV}$ for $x_P=0.1$, $y=0.5$, $|t|=0.3 \mbox{GeV}^2$:
dotted line-for $Q^2=0.5 \mbox{GeV}^2$; solid line-for $Q^2=0.5 \mbox{GeV}^2$;
dot-dashed line-for $Q^2=5 \mbox{GeV}^2$; dashed line-for $Q^2=10 \mbox{GeV}^2$.}
\label{altlqs}       
\end{figure}
\begin{figure}
\epsfxsize=10,8cm \centerline{\epsfbox{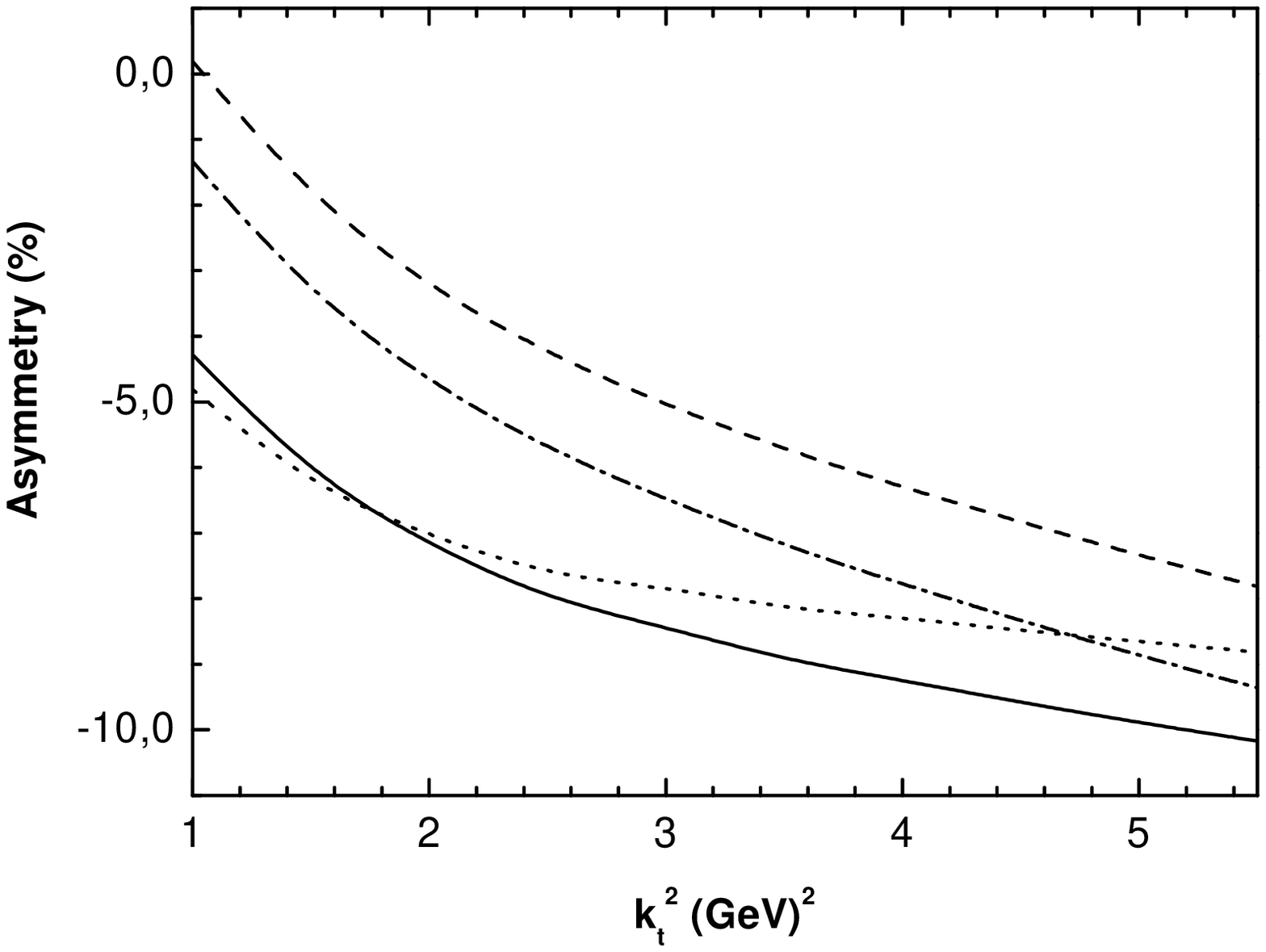}}
\caption{The
$A^Q_{lT}$ asymmetry in diffractive heavy $Q \bar Q$ production at
$\sqrt{s}=20 \mbox{GeV}$ for $x_P=0.1$, $y=0.5$, $|t|=0.3
\mbox{GeV}^2$: dotted line-for $Q^2=0.5 \mbox{GeV}^2$; solid
line-for $Q^2=0.5 \mbox{GeV}^2$; dot-dashed line-for $Q^2=5
\mbox{GeV}^2$; dashed line-for $Q^2=10 \mbox{GeV}^2$.}
\label{althqs}       
\end{figure}
\begin{figure}
\epsfxsize=10,8cm \centerline{\epsfbox{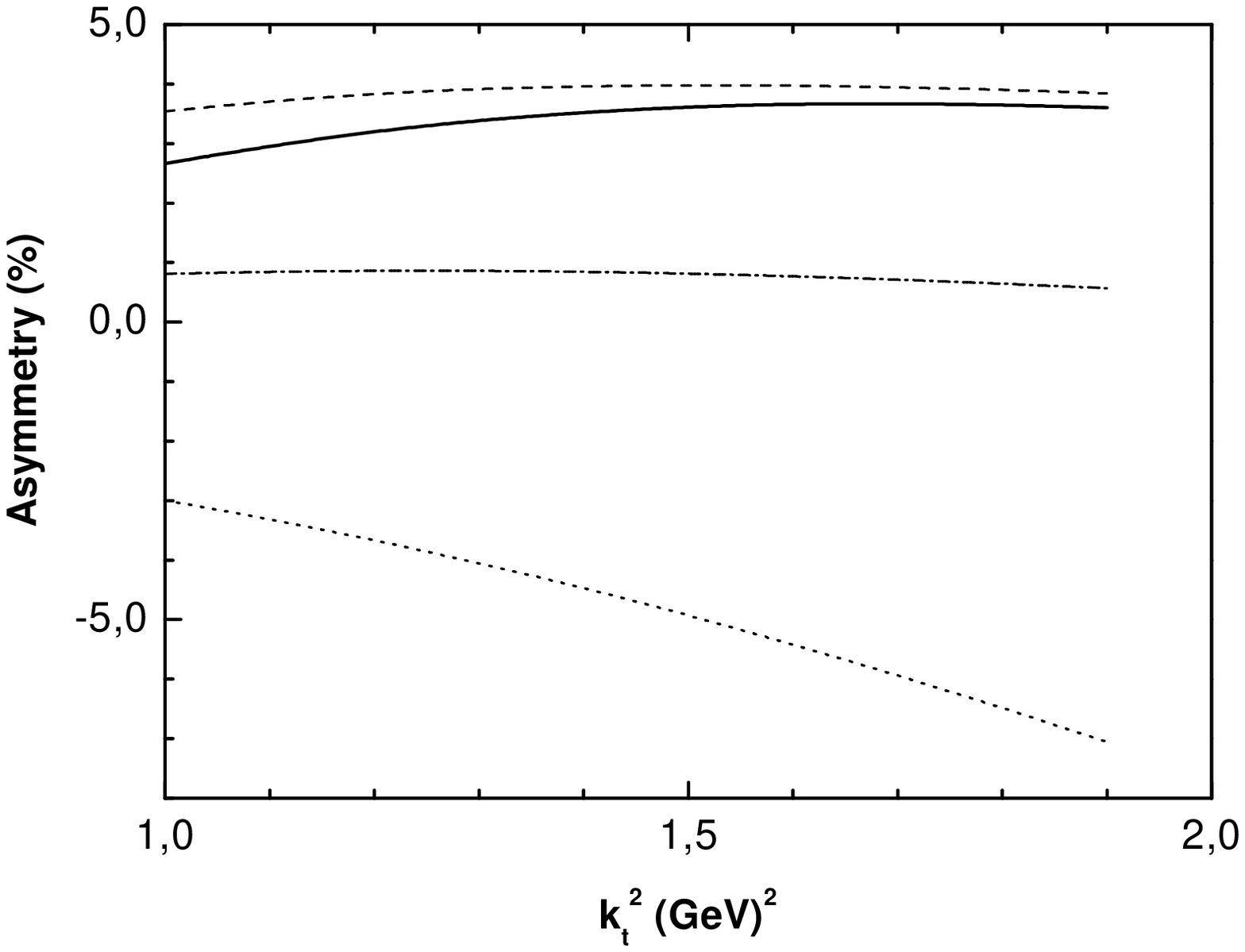}} \caption{The
The $A^Q_{lT}$ asymmetry  in diffractive light $Q \bar Q$
production at  $\sqrt{s}=7 \mbox{GeV}$ for $Q^2=5\mbox{GeV}^2$,
$x_P=0.1$, $y=0.5$: dotted line -at $|t|=0.1 \mbox{GeV}^2$;
dot-dashed line -at $|t|=0.3 \mbox{GeV}^2$; dashed line -at
$|t|=0.5 \mbox{GeV}^2$; solid line- integrated over $|t|$
asymmetry.}
\label{altl7qs}       
\end{figure}
At  the  energy $\sqrt{s} =7 \mbox{GeV}$ (HERMES) it is not so
easy to study the perturbative region for $Q \bar Q$ production.
Really, $k_\perp^2$ should be large enough to have a large scale
$k_0^2$ in the process (\ref{kqq}). Otherwise, from (\ref{int}),
we have the restriction that $k^2\le M_X^2/4$. In this energy
range for quite large $M_X^2 \sim (8-10) \mbox{GeV}^2 \sim
M^2_{J/\Psi}$ we find that $(k_\perp^2)_{max} \sim 2\mbox{GeV}^2$.
This means that we can work only in a very limited region of
$k^2$. The expected $A_{lT}$ asymmetry for light quark production
at HERMES is shown in Fig.\ref{althm}. We find that for
$k_\perp^2=1.3\mbox{GeV}^2$, $Q^2=5\mbox{GeV}^2$, $x_P=0.1$,
$y=0.5$, and $|t|=0.3 \mbox{GeV}^2$, the coefficient $C_k^{Q \bar
Q}$ in (\ref{cltqq}) is quite large, about 1.5 at the HERMES
energy. This shows a possibility to study the polarized gluon
distribution ${\cal K}^g_\zeta(x)$ in the HERMES experiment.

The contribution to asymmetry $\propto \vec Q \vec S_\perp$ is
simpler to study experimentally. Moreover, this term is connected
directly with the diffractive contribution with the $A_{\perp}$
asymmetry \cite{efrem}. We shall analyze this term for the case
when the transverse jet momentum $\vec Q_\perp$ is parallel to the
target polarization $\vec S_\perp$ (a maximal contribution to the
asymmetry). The predicted $A^Q_{lT}$ asymmetry in diffractive
light $Q \bar Q$ production at $\sqrt{s} =20 \mbox{GeV}$ is shown
in Fig. \ref{altlqs}. This asymmetry is not small for $Q^2 \sim
(0.5-1) \mbox{GeV}^2$. In contrast to the $A^k_{lT}$ term, the
$A^Q_{lT}$ asymmetry has a strong mass dependence. For heavy quark
production, this asymmetry becomes negative, Fig. \ref{althqs}.

It is interesting to look for what we expect to observe for light
quark production at low energy $\sqrt{s} =7 \mbox{GeV}$. The
predicted asymmetry for different momentum transfers is shown in
Fig. \ref{altl7qs}. Note that in fixed--target experiments, it is
usually difficult to detect the final hadron and determine the
momentum transfer. In this case, it will be good to have
predictions for the asymmetry integrated over momentum transfer
\begin{equation}\label{intasy}
\bar A^Q_{lT}= \frac{\int_{t_{min}}^{t_{max}}\,\sigma(-)\,dt}
{\int_{t_{min}}^{t_{max}}\,\sigma(+)\,dt}.
\end{equation}
We integrate cross sections from $t_{min} \sim (x_P m)^2 \sim 0$
up to $t_{max}=4 \mbox{GeV}^2$. The predicted integrated asymmetry
(see Fig. \ref{altl7qs}) is not small, about 3\%.

\section{Conclusion}\label{sect8}
In the present paper, diffractive hadron leptoproduction for a
longitudinally polarized lepton and a transversely polarized
proton at high energies has been studied within the two-gluon
exchange model. The polarized cross sections of diffractive hadron
production are determined in terms of the leptonic and hadronic
tensors and the squared amplitude of hadron production through the
photon-two-gluon fusion. The hadronic tensor is expressed in terms
of the two--gluon couplings with the proton that are related to
SPD. As a result, the cross sections of diffractive meson and $Q
\bar Q$ production are expressed in terms of the same integrals
which are connected with the gluon SPD ${\cal F}_\zeta(x)$ and
${\cal K}_\zeta(x)$.

The $A_{lT}$ asymmetry is found to be proportional to the ratio of
${\cal K}/{\cal F}$ structure functions and generally can be used
to get information on the transverse distribution ${\cal
K}^g_{x_P}(x_P,t)$ from experiment.  The asymmetry of vector meson
production is expected to be quite small, $A_{lT}< 0.1 \%$ in the
HERMES energy range. This result was obtained for a simple
nonrelativistic form of the vector meson wave function and
generally can be used only for heavy meson production.  The
$A_{lT}$ asymmetry for $Q \bar Q$ production contains  two
independent terms which are proportional to the scalar products
$\vec k_\perp \vec S_\perp$ and $\vec Q \vec S_\perp$ (\ref{nm}).
The first one $\propto \vec k_\perp \vec S_\perp$ have no $x_P$
suppression and is predicted to be about 10\%.  It might be an
excellent object to study transverse effects in the proton
coupling with gluons. However, the experimental study of this
asymmetry is not so simple. To find nonzero asymmetry in this
case, it is necessary to distinguish quark and antiquark jets and
to have a possibility to study the azimuthal asymmetry structure.
This is an important condition, because the polarized cross
section integrated over $d \phi_{Jet}$ is equal to zero. Note that
the asymmetry of the same order of magnitude was predicted for
diffractive $Q \bar Q$ production in polarized proton- proton
interaction \cite{golos96}.

The second term $\propto \vec Q \vec S_\perp$ in the $A_{lT}$
asymmetry of  $Q \bar Q$ production is related to the diffractive
contribution to the $A_\perp$ asymmetry. The expected asymmetry in
this case is not small too. The predicted coefficient $C_Q^{Q \bar
Q}$  in (\ref{cltqq}) is about  0.3. This shows the possibility to
study the ratio of polarized gluon distributions ${\cal K}/{\cal
F}$ in diffractive  $Q \bar Q$ leptoproduction. All these results
should be applicable to the reactions with heavy quarks. For
processes with light quarks, our predictions can be used in the
small $x$ region ($x \le 0.1$ e.g.) where the contribution of
quark SPD is expected to be small.

Now, we have no the definite predictions for the $A_{{lT}}$
asymmetry in light vector meson production. Including the
transverse quark motion and higher twist effects for transversely
polarized $\rho$ meson might be important in asymmetry. In the
region of not small $x\ge 0.1$ in the HERMES experiments, the
polarized quark SPD might be studied together with the gluon
distribution in the case of  $\rho$ production. In the case of
$\phi$ production, the strange quark SPD might be analyzed.
Similar experiments can be conducted at the future COMPASS
spectrometer if a transversely polarized target is constructed
there.

We conclude that important information on the spin--dependent SPD
at small $x$ can be obtained from the asymmetries in diffractive
hadron leptoproduction for longitudinally polarized lepton and
transversely polarized hadron targets.

We would like to thank  A. Borissov, A. Efremov, P. Kroll, T.
Morii, O. Nachtmann, W.-D. Nowak, and O. Teryaev for fruitful
discussions.\\[0.2cm]

This work was supported in part by the Russian Foundation for
Basic Research, Grant 00-02-16696.\\[1cm]



\begin{thebibliography}{999}
\bibitem{rad} A.V. Radyushkin, Phys.Rev, D {\bf 56}, 5524 (1997).
\bibitem{ji} X. Ji, Phys.Rev. D {\bf 55} 7114 (1997).
\bibitem{coll} J.C. Collins, L. Frenkfurt, M. Strikman, Phys.\ Rev.\
               {\bf D56},  2982 (1997).
\bibitem{low} F.E.\ Low, Phys.\ Rev.\ {\bf D12},163  (1975) \\
              S.\ Nussinov, Phys.\ Rev.\ Lett.\ {\bf 34}, 1286 (1975).
\bibitem{zeus97} ZEUS Collab., J. Breitweg et al., Z. Phys,
{\bf C75}, 215 (1997).
\bibitem{jpsi1} H1 Collaboration,  S. Aid et al., Nucl. Phys. {\bf B472},
 3 (1996).
\bibitem{h1_99} H1 Collab., C. Adloff et al., Eur. Phys. J. {\bf C10},  373 (1999).
\bibitem{dijet} ZEUS Collab., J. Breitweg et al., Eur. Phys. J. {\bf C5}, 41 (1998);\\
               H1 Collab., C. Adloff et al., Eur. Phys. J. {\bf C6}  421 (1999).
\bibitem{hermes} HERMES Collab., A. Airapetian et al, Phys. Lett. {\bf B513}, 301 (2001).
\bibitem{rys} M.G. Ryskin, Z. Phys. {\bf C57}, 89 (1993).
\bibitem{J-Psi} S.J. Brodsky at al., Phys. Rev. {\bf D50}, 3134 (1994).
\bibitem{clebaux}B. Clerbaux, Elastic production of Vector Mesons at HERA:
  study of the scale of the interaction and measurement of the helicity amplitudes.
  E-print: hep-ph/9908519.
\bibitem{rys1} M.G. Ryskin, R.G. Roberts, A.D. Martin, E.M. Levin,
     Z. Phys.  {\bf C76}, 231 (1997).
\bibitem{cud} J.L. Cudell, I. Royen, Nucl. Phys. {\bf B545}, 505 (1999).
\bibitem{ivan} D.Y. Ivanov, R. Kirshner,  Phys. Rev. {\bf D58}, 114026 (1998).
\bibitem{mank98} L.\ Mankiewicz, G. Piller, T. Weigl,  Eur. Phys. J. {\bf C5}, 119 (1998).
\bibitem{hua_kr} H.W. Huang, P. Kroll, Eur. Phys. J. {\bf C17},  423 (2000).
\bibitem{mank} M.\ V\"anttinen, L.\ Mankiewicz, Phys.Lett.  {\bf B434}, 141 (1998).
\bibitem{mank00} L.\ Mankiewicz, G. Piller, Phys. Rev. {\bf D61}, 074013 (2000).
\bibitem{die95} M.Diehl, Z. Phys. {\bf C66}, 181 (1995).
\bibitem{bart96} J. Bartels, C. Ewerz, H. Lotter, M.W\"usthoff,
                Phys. Lett. {\bf B386}, 389 (1996).
\bibitem{ryskin97}  E.M. Levin, A.D. Martin, M.G. Ryskin, T. Teubner,
            Z. Phys.  {\bf C74}, 671 (1997).
\bibitem{schaef} B. Lehmann-Dronke, M. Maul, S. Schaefer, E.Stein, A. Sch\"afer,
       Phys.Lett. {\bf B457}, 207 (1999).
\bibitem{bart99} J. Bartels, T. Gehrmann, M.G. Ryskin, Eur. Phys. J. {\bf C11}, 325 (1999).
\bibitem{transverse} Proceedings of the Topical Workshop
    "Transverse Spin Physics", DESY Zeuthen, July 2001, editted by:
    J. Bl\"umlein, W.-D. Nowak, G. Schnell, Internal Report DESY
    Zeuthen 01-01 August 2001.
\bibitem{golos01} S.V.\ Goloskokov, Spin effects in diffractive hadron
photoproduction.   Proc. of the 14th International Spin Physics Symposium,
SPIN2000,  editted by: K. Hatanaka, T. Nakano, K. Imai, H. Ejiri.
AIP Conference  Proc. {\bf V.570}, 541; hep-ph/0011341; hep-ph/0110212.
\bibitem{efrem}  M. Anselmino, A. Efremov, E. Leader, Phys.
               Rept. {\bf 261}, 1 (1995).
\bibitem{kroll} M.\ Anselmino, P.\ Kroll, B.\ Pire,
                Z. Phys. {\bf C36}, 36 (1987).
\bibitem{grib77} L.V. Gribov, E.M. Levin,  M.G. Ryskin, Phys.
             Rept. {\bf 100}, 151 (1983).
\bibitem{gol_kr} S.V.\ Goloskokov, P.\ Kroll, Phys.\ Rev.
         D {\bf 60}, 014019 (1999).
\bibitem{gol93} S.V.\ Goloskokov,  Phys.Lett. {\bf B315}, 459 (1993).
\bibitem{gol_mod}  S.V. Goloskokov, S.P. Kuleshov, O.V. Selyugin,
           Z. Phys. {\bf C50},  455 (1991).
\bibitem{krish} D.C. Peaslee et al., Phys.\ Rev.\ Lett.\ {\bf 51}, 2359 (1983).
\bibitem{akch}  N. Akchurin, S.V. Goloskokov, O.V. Selyugin,
Int.J.Mod.Phys. {\bf A14}, 253 (1999).
\bibitem{berger} E.L. Berger, D. Jones, Phys. Rev. {\bf D23}, 1521 (1981).
\bibitem{goljpsi} S.V. Goloskokov,
 Eur. Phys. J. {\bf C11}, 309 (1999).
 \bibitem{golwf} S.V. Goloskokov, On the $\sigma _L/\sigma _T$ ratio in polarized
vector meson  photoproduction. Proc. of XV
International Seminar on High Energy Physics Problems
"Relativistic Nuclear Physics and
  Quantum Chromodynamics", Dubna September 25-29, 2000; hep-ph/0012307.
\bibitem{lansh-m} A.\ Donnachie, P.V.\ Landshoff,
                  Nucl.\ Phys.\ {\bf B244}, 322 (1984).
\bibitem{nach} T. Arens, M. Diehl, O. Nachtmann, P.V. Landshoff,
     Z. Phys. {\bf C74}, 651 (1997).
\bibitem{zeus_p} ZEUS Collab., J. Breitweg et al.,
       Eur. Phys. J. {\bf C14}, 213 (2000).
\bibitem{golos96} S.V.Goloskokov, Phys.Rev. {\bf D53}, 5995 (1996).
\end{thebibliography}
\end{document}